\documentclass[%
 aip,
 amsmath,amssymb,
preprint,%
]{revtex4-1}

\usepackage{graphicx}
\usepackage{dcolumn}
\usepackage{bm}
\usepackage{nicefrac}

\usepackage[utf8]{inputenc}
\usepackage[T1]{fontenc}
\usepackage{mathptmx}
\usepackage{etoolbox}
\usepackage[dvipsnames]{xcolor}
\usepackage{titlesec}
\makeatletter
\def\@email#1#2{%
 \endgroup
 \patchcmd{\titleblock@produce}
  {\frontmatter@RRAPformat}
  {\frontmatter@RRAPformat{\produce@RRAP{*#1\href{mailto:#2}{#2}}}\frontmatter@RRAPformat}
  {}{}
}%
\makeatother

\begin{document}
\preprint{AIP/123-QED}
\title{A review of Girsanov Reweighting and of Square Root Approximation for building molecular Markov State Models}
\author{Luca Donati}
\author{Marcus Weber}
\author{Bettina G. Keller}

\date{\today}


\begin{abstract}
Dynamical reweighting methods permit to estimate kinetic observables of a stochastic process governed by a target potential $\tilde{V}(x)$ from trajectories that have been generated at a different potential $V(x)$. 
In this article, we present Girsanov reweighting and Square Root Approximation (SqRA): the first method reweights path probabilities exploiting the Girsanov theorem and can be applied to Markov State Models (MSMs) to reweight transition probabilities; the second method was originally developed to discretize the Fokker-Planck operator into a transition rate matrix, but here we implement it into a reweighting scheme for transition rates.
We begin by reviewing the theoretical background of the methods, then present two applications relevant to Molecular Dynamics (MD), highlighting their strengths and weaknesses.
\end{abstract}


\maketitle
\section{Introduction}
Molecular Dynamics (MD) simulations permit to investigate the conformational ensemble and the dynamics of molecular systems calculating the inter-atomic forces and integrating the associated equations of motion \cite{Haile1997, Frenkel2002}, producing time-discretized trajectories $\omega = \lbrace x_0,...,x_n\rbrace$ which contain the position of the atoms at each time-step.
However, solving the equations of motion of a high dimensional system requires huge computational resources due to the high potential energy barriers that prevent the exploration of the conformational space of the system; thus, even the most efficient computers need several months to simulate rare events, such as the opening of a closed ligand-bound conformation, that occur at milliseconds timescales \cite{Shaw2008, Shaw2009, Ramanathan2009, Ramanathan2014}. 

In the last years, the research in this field focused on the reduction of the computational cost, and on the optimization of the post-analysis of MD trajectories, developing varieties of methods and algorithms.
For example, enhanced sampling methods such as Umbrella sampling \cite{Souaille2001} and metadynamics \cite{Huber1994, Laio2002}, perform MD simulations at a biased potential $\tilde{V}(x) = V(x) + U(s)$, where the bias $U(s)$ reduces the height of the potential energy barriers along  the reaction coordinates $s$ that well describe the kinetic properties in a low-dimensional subspace of the system; while the replica exchange method \cite{Sugita2000} performs parallel simulations at different temperatures to facilitate the jump between metastable states.
Transition Path Sampling (TPS) \cite{Bolhuis2002, Dellago2002} samples the transition channel of the system, generating an ensemble of unbiased trajectories via a Metropolis Monte Carlo procedure.
Post-processing techniques such as Markov State Models (MSMs) \cite{Schuette1999, Schuette1999b,  Deuflhard2000, Swope2004, Chodera2007, Buchete2008, Prinz2011, Keller2010, Keller2011} and variational methods \cite{Nuske2014, Nuske2016, Vitalini2015b}, make use of MD trajectories and time-correlation functions to build a transition probability matrix $\mathbf{T}(\tau)$, discretization of the transfer operator $\mathcal{T}(\tau)$, that describes the high-dimensional dynamics as a stochastic process on few relevant coordinates.
However, these classes of methods suffer from two problems.
On the one side, enhanced sampling simulations improve the sampling, but change drastically the dynamics of the system.
Furthermore, they require prior knowledge of the
reaction coordinates to perturb.
TPS methods yield the unbiased dynamics of the system and permit to extract the reaction coordinates, but they are computationally expensive and limited to two-states systems.
MSMs require trajectories that exhaustively sample the ensemble, but that are generated by unbiased simulations.

Dynamical reweighting methods permit to solve these issues and to recover the correct dynamical properties of the system from biased simulations.
For example, the Transition-Based Reweighting Analysis Method (TRAM) \cite{Wu2014, Wu2016} assumes that the dynamics is in local equilibrium within each subset of the state space and the MSM transition probabilities are reweighted using a maximum likelihood estimator.
In parallel tempering simulations the path probability density of a time discretized path, generated at a reference temperature, is reweighted to the target temperature to build a MSM \cite{Chodera:2011, Prinz:2011b}.
A different reweighting approach exploits a stochastic path integral formulation of the problem and the use of the Onsager-Machlup (OM) action for weighting continuous trajectories generated by stochastic processes \cite{Onsager1953}.
To be more precise, the OM action serves to estimate the unnormalized probability that a stochastic process generates trajectories lying within a small tubular neighborhood of a smooth path \cite{Hartmann2013}, or alternatively it can be interpreted as the Lagrangian giving the most probable tube within which a trajectory can be found \cite{Durr1978}.
Most notable to mention is the work by Zuckerman and Woolf \cite{Zuckerman2000}, where the OM action is used to optimize a dynamic importance sampling (DIMS) method \cite{Woolf1998, Zuckerman1999} for the calculation of transition rates.
Nonetheless, in such studies, the OM action is not employed for dynamical reweighting, but to determine the most probable crossing event and to recursively correct the bias during the simulation in order to increase the efficiency of sampling.
Instead, in a more recent work by Xing and Andricioaei \cite{Xing2006}, the OM action was implemented in a reweighting scheme, where the ratio between path weights is used to estimate time-correlation functions from trajectories generated by scaled potentials to reduce the height of the barriers.
However this approach is limited by the lag time of the time-correlation function.
Indeed, the path integral of the weights increases dramatically with time, producing noisy tails of the time-correlation functions.

In 2017, we proposed a path reweighting method, referred to as Girsanov reweighting, based on the Girsanov theorem \cite{girsanov1960}, an important result from stochastic analysis \cite{Oksendal2003} that guarantees the existence of the ratio between probability measures under the condition of absolute continuity of the measures.
The Girsanov theorem can be applied also to path measures, and, despite being chronologically derived later, it provides the theoretical formalism within the path probabilities based on the OM action can be explicitly derived \cite{Durr1978}.
We applied the Girsanov reweighting procedure to MSMs \cite{Schuette2015b, Donati2017}, multiplying the time-correlation functions between subsets of the state space, by the ratio of path probabilities associated to a target and a reference system.
In 2018 we further expanded the method and we applied it to metadynamics simulations \cite{Donati2018}, where the potential of a molecular system is perturbed by a cumulant sum of Gaussian functions, deposited along the relevant coordinates during the simulation.
In a second simulation, the bias thus constructed was used for a quick sampling of the ensemble, and the trajectory was reweighted to build the MSM of the unbiased system.
This strategy solved the problems encountered by Xing and Andricioaei. 
Indeed, by considering a time- and space-discretized model of the dynamics (i.e.~an MSM), the lag-time of the reweighted correlation functions is limited by the time-step $\tau$ of the model (i.e.~the MSM lag time). 
This lag time $\tau$ is considerably (often: order of magnitude) shorter than the timescales of the rare events in the system. 
We can thus avoid the numerical problems that arise when reweighting correlation functions with long lag times. 
In addition, Metadynamics perturbs only few reaction coordinates, leaving unbiased most of the force field terms.

In this work, we review the underlying theory of Girsanov reweighting and we propose new applications.
In a first example, we studied a convex combination of two potentials $V_A(x)$ and $V_B(x)$, interpolated by a $\lambda$-potential $V(x;\lambda) = \lambda V_A(x) + (1-\lambda) V_B(x)$, with $\lambda \in [0,1]$.
Given a simulation, carried out at an appropriate $\lambda$ value, we can build the MSM for all the other $\lambda$ values.
In the second example, we applied Girsanov reweighting to a 6-atom prototype molecule for force field parametrization, i.e. to study the sensitivity of the system as a function of its parameters.

In the second part, we also review the theory of the Square Root Approximation of the Fokker-Planck operator \cite{Lie2013, Donati2018b, Heida2018, Donati2021}.
This method permits to build the rate matrix $\mathbf{Q}$ of a stochastic process, whose entries, the rates between the subsets of the state space, are estimated as geometrical averages of their stationary weights.
Since the transition rate matrix $\mathbf{Q}$ and the transition probability matrix $\mathbf{T}(\tau)$ are strictly related and share the same eigenspace, SqRA and MSMs can be used alternatively to describe the dynamics of a molecular system as a stochastic process. 
The main advantage of SqRA is that it does not rely on time-correlation functions, but makes an efficient use of stationary distributions.
This is particularly useful for low-dimensional systems, as it yields the exact discretization of the Fokker-Planck operator with no need of running  simulations.
For this reason, in the first example, we used the SqRA to validate the results obtained by path reweighting.
In addition, in this work, we propose a new dynamical reweighting scheme based on SqRA, where a rate matrix $\mathbf{Q}$ associated to a reference potential $V(x)$ is transformed into a rate matrix $\tilde{\mathbf{Q}}$ of a target potential $\tilde{V}(x)$. 
A similar strategy\cite{Bicout1998} was already implemented by Rosta and Hummer \cite{Rosta2014} into a dynamic histogram analysis method (DHAM) for umbrella sampling simulations to reweight transition probabilities with small lag time, while here we apply the method to transition rates.
We tested this new method in our second example, comparing the results with those obtained by the path reweighting procedure.

The article is outlined as follows.
\begin{enumerate}
    \item In section \ref{sec:girsanov}, we review the theory of the path reweighting method based on the Girsanov theorem for Brownian dynamics.
    \item In section \ref{sec:sqra}, we introduce SqRA method to discretize the infinitesimal generator and we propose a reweighting scheme for rate matrices.
    \item In section \ref{sec:resA}, we study a convex combination between two two-dimensional potentials based on the M\"uller-Brown Potential \cite{Muller1979}.
    \item In section \ref{sec:resB}, we study the effect of the electric constant in a 6-atoms prototype molecule where non-bonded atoms interact through the Coulomb potential.
    \item In section \ref{sec:disc_sec}, we discuss the results and draw conclusions, highlighting the advantages and the limitations of both the methods.
\end{enumerate}
\begin{figure*}[ht]
\includegraphics[width=1.0\textwidth]{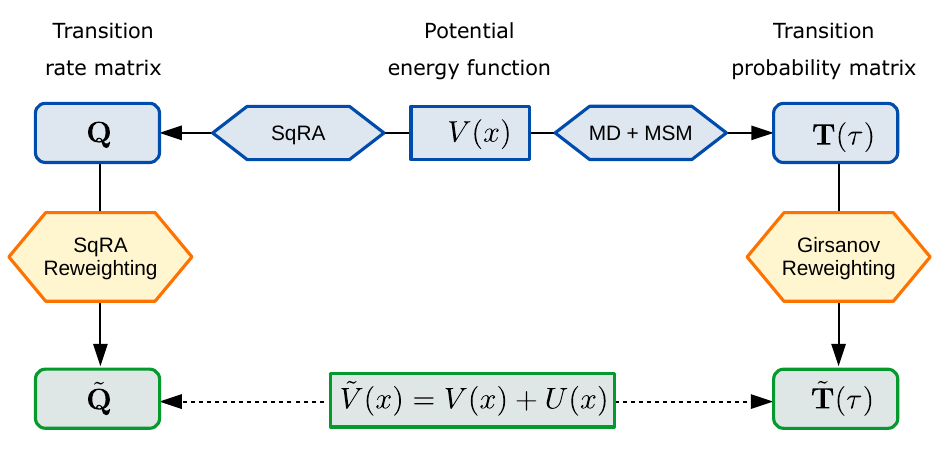}
\caption{Workflow of the methods discussed in the article. Given a dynamical system with potential energy function $V(x)$, the rate matrix $\mathbf{Q}$ is constructed via \emph{SqRA} (eq.~\ref{eq:finalRateMatrix}), the transition probability matrix $\mathbf{T}(\tau)$ is built via \emph{MSM} from an MD simulation.
Given a perturbed potential $\tilde{V(x)} = V(x) + U(x)$, the rate matrix $\tilde{\mathbf{Q}}$ and the probability matrix $\tilde{\mathbf{T}}(\tau)$ are constructed reweighting respectively the matrix $\mathbf{Q}$ via \emph{SqRA reweighting} (eq.~\ref{eq:sqraReweighting}), and the matrix $\mathbf{T}(\tau)$ via \emph{Girsanov reweighting} (eq.~\ref{eq:correlationReweighted}).}
\label{fig:fig1}
\end{figure*}

\section{Girsanov theorem and path reweighting}
\label{sec:girsanov}
\subsection{Brownian dynamics}
Consider a molecular system of $N_a$ atoms that move in the three-dimensional Cartesian space $\mathbb{R}^{3}$.
The \emph{state space} of the system is denoted by 
$\Gamma \subset \mathbb{R}^{N_D}$, 
where $N_D = 3 N_a$ is the total number of dimensions.
Let the system be governed by the overdamped Langevin dynamics, which is described by the stochastic differential equation
\begin{equation}
    \mathrm{d}x(t) = -\,  \boldsymbol{\xi}^{-1} \mathbf{M}^{-1} \, \nabla V(x(t)) \mathrm{d}t + \boldsymbol{\sigma} \mathrm{d} W(t) \, ,
   \label{eq:sde}
\end{equation}
where $x(t) \in \Gamma$ is the state vector at time $t$, 
$\boldsymbol{\xi}$ and $\mathbf{M}$ are respectively two $3 N_a \times 3 N_a$ diagonal matrices containing the friction and mass of each atom for each direction, $V(x)$ is the potential energy function,
and $W(t)$ is an $N_D$-dimensional Wiener process
scaled by the diagonal matrix $\boldsymbol{\sigma} = \sqrt{2 k_B T \boldsymbol{\xi}^{-1}  \mathbf{M}^{-1}} = \sqrt{2 \mathbf{D}}$ 
where $T$ is the temperature $k_B$ is the Boltzmann constant and  $\mathbf{D} = \nicefrac{1}{2} \boldsymbol{\sigma}^2$ is the diffusion matrix.
Eq.~\ref{eq:sde} generates a Markovian, ergodic and reversible process \cite{Schuette1999b, Risken1989}. 
%

\subsection{Path space}
Let $\omega=\lbrace x_0= x, \,  x_1,. ...,\,  x_n \rbrace$ be a time-discretized \emph{path} of length $\tau = n \cdot \Delta t$, starting at $x_0 = x$.
Suppose that $\omega$ is an approximated solution of eq.~\ref{eq:sde} generated using the Euler-Maruyama scheme using an integration time step $\Delta t$ and a sequence of independent and identically distributed random numbers $\eta_{k}^i$ drawn from a Gaussian distribution generated at each timestep $k$ for each dimension $i$ of the system.
The path $\omega$ is an element of the \emph{path space}, i.e. the set $\Omega_{\tau,x} = \Gamma^n \subset \mathbb{R}^{3N_a \cdot n}$ that contains all the possible paths.

The \emph{path probability density} of the path $\omega \in \Omega_{\tau,  x}$ is the product of the conditional probabilities between consecutive steps:
\begin{equation}
\begin{aligned}
\mu_P(\omega) & =  \mu_P( x_1,  x_2, ... ,  x_n \, |\,  x_0 =  x) \\
& =  p( x_0,  x_1; \Delta t) \cdot  p( x_1,  x_2; \Delta t) \cdot ... \cdot p( x_{n-1},  x_n; \Delta t) \, ,
\label{eq:path_prob_dens}
\end{aligned}
\end{equation}
where the conditional probability that the system visits the state $x_{k+1}$, given that the previous state was $x_k$, is
\begin{eqnarray}
\label{eq:ConditionalProbability}
p( x_{k-1},  x_k; \Delta t)
& = &
\mathcal{N}\,\exp \left\lbrace - \frac{1}{2 \Delta t} \times \right.\cr
&& \times \left. \left[x_{k} - x_{k-1} - \nabla V (x_{k-1})\right]^\top \times \right.\cr
&& \times \left.\boldsymbol{\sigma}^{-2} \times \right.\cr
&& \times \left.\left[x_{k} - x_{k-1} - \nabla V (x_{k-1})\right]
\right\rbrace \, .
\end{eqnarray}
In eq.~\ref{eq:ConditionalProbability}, we introduced the normalization constant 
\begin{eqnarray}
\mathcal{N} 
&=& 
\prod_{i=1}^{3N_a} 
\left(\frac{1}{{2\pi \Delta t \sigma_i^2}}\right)^{\frac{1}{2}} \cr
&=& 
\frac{1}{\det{\left({2\pi \Delta t \, \boldsymbol{\sigma}}^2\right)}^{\frac{1}{2}}} \, ,
\end{eqnarray}
where $\sigma_i^2$ is the $i$th diagonal element, one for each direction of each atom, of the $3N_a \times 3N_a$ diagonal matrix $\boldsymbol{\sigma}^2$.

Let us restrict the domain to a \emph{subset of the path space} $\mathcal{A}$, which is constructed as a product of subsets $A_i \subset \Gamma$ of the state space
$\mathcal{A}=A_1 \times A_2 ... \times A_n$.
Each set $A_i$ represents a region of the state space in which $x_i$ may be found.
The associated \emph{path probability measure} is the integral
\begin{eqnarray}
\label{eq:path_measure1}
P(\mathcal{A})
	&= &  \mathbb{P}(\omega \in \mathcal{A}) = \mathbb{P}(x_1 \in A_1,x_2 \in A_2,...,x_\tau \in A_n) \cr
	& = & \int_\mathcal{A}\mu_P(\omega) \, \mathrm{d}\omega \cr
	& = & \int_{A_1}\int_{A_2}...\int_{A_n} 
			p(x_0,x_1; \, \Delta t) \, p(x_1,x_2; \, \Delta t)\, ... \cr
	&& ... \, p(x_{n-1},x_n;\, \Delta t) 	  \, \mathrm{d}x_1 \, \mathrm{d}x_2 \, ... \, \mathrm{d}x_n \, ,
\end{eqnarray}
which describes the probability to find the path $\omega$ in the subset $\mathcal{A}$ of the path space.

\subsection{Path reweighting}
Consider now a perturbed potential energy function 
\begin{eqnarray}
    \tilde{V}(x) = V(x) + U(x) \, .
    \label{eq:perturbedPotential}
\end{eqnarray}
The paths of the path space $\Omega_{\tau, x}$ are still acceptable solutions of eq.~\ref{eq:sde}, but the associated path probability density $\mu_{\tilde{P}}(\omega)$ changes according to eq.~\ref{eq:ConditionalProbability}, inducing also a modification of the probability that the path $\omega$ belongs to the same subset $\mathcal{A}$ of the path space
\begin{eqnarray}
    \tilde{P}(\mathcal{A})
	&= &  \mathbb{\tilde{P}}(\omega \in \mathcal{A}) = \mathbb{\tilde{P}}(x_1 \in A_1,x_2 \in A_2,...,x_\tau \in A_n) \cr
	& = & \int_\mathcal{A}\mu_{\tilde{P}}(\omega) \, \mathrm{d}\omega \, .
\end{eqnarray}

How are the probability densities $\mu_{{P}}(\omega)$ and $\mu_{\widetilde{P}}(\omega)$ related? 
The \emph{Radon-Nikodym theorem} \cite{Rudin1986, Oksendal2003} asserts that if the condition
\begin{eqnarray}
\widetilde P(\mathcal{A}) &&= \int_\mathcal{A}  \mu_{\widetilde P}(\omega)  \mathrm{d}\omega=0  \Rightarrow  \cr
&& \Rightarrow
P(\mathcal{A}) = \int_\mathcal{A} \mu_{P}(\omega) \mathrm{d}\omega= 0 \qquad
\forall \mathcal{A} \subset \Omega_{\tau, x}\, 
\label{eq:absContPath}
\end{eqnarray}
holds, then there exists the \emph{Radon-Nikodym derivative}
\begin{eqnarray}
    M_{\tau,  x}(\omega) = \frac{\mu_{\tilde{P}}(\omega)}{\mu_P(\omega)}
\label{eq:RadonNikodym}  
\end{eqnarray}
and for any measurable set $\mathcal{A}\in \Omega_{\tau,x}$
\begin{eqnarray}
\tilde{P}(\mathcal{A}) = \int_\mathcal{A} M_{\tau,  x}(\omega) \, \mu_P(\omega) \, \mathrm{d}\omega 
\end{eqnarray}
The condition expressed by eq.~\ref{eq:absContPath} is known as \emph{absolute continuity} of the measure $\tilde{P}$ with respect to the measure $P$.
In literature this property is denoted by $\tilde{P} \ll P$.
Intuitively, eq.~\ref{eq:absContPath} means that any region of the path space $\mathcal{A}$ that is sampled by the dynamics at $\tilde V(x)$, also needs to be sampled by the dynamic at $V(x)$, 
i.e.~``the path probability measures need to overlap''.
If this is not the case, the relative path probability density in eq.~\ref{eq:RadonNikodym} is not defined.

For stochastic processes governed by the overdamped Langevin dynamics equation, the ratio between path probability densities reads
\begin{align}
&M_{\tau,  x}(\omega) = \frac{\mu_{\tilde{P}}(\omega)}{\mu_P (\omega)} =   \cr
&\exp \left\lbrace  \sum_{i=1}^{3N_p} \left[
\sum_{k=0}^{n-1}
- \frac{\nabla_i U(x_k)}{\sigma_i} 
\eta_k^i 
\sqrt{\Delta t}
- \frac{1}{2} 
\sum_{k=0}^{n-1}
\left(\frac{\nabla_i U(x_k)}{\sigma_i} \right)^2 \Delta t 
\right] \right\rbrace \, , \cr
\label{eq:girsanov}
\end{align}
where $\eta_k^i$ are the random numbers generated during the simulation to solve eq.~\ref{eq:sde}, and the expression $\eta_k^i \sqrt{\Delta t}$ is used to approximate the stochastic integral according to the Ito convention.
In appendix \ref{sec:appendix1}, we report a derivation of eq.~\ref{eq:girsanov} for one-dimensional processes.
Note that eq.~\ref{eq:girsanov} is derived for overdamped Langevin dynamics applying the Euler-Maruyama scheme, but different expressions can be derived for different dynamics and for different numerical integrators.
For a detailed discussion, see Ref.~\onlinecite{Kieninger2021},  where additionally an exact expression for time-discretized trajectories generated by underdamped Langevin dynamics with a simple Langevin integrator is derived \cite{Eastman2013}. 
\subsection{Path ensemble average}
Let $f(\omega) = f( x_1,  x_2, ... ,  x_n)$ be a \emph{path observable}, i.e. a suitable function which assigns a real-valued number to each path $\omega$, then the \emph{path ensemble average} of $f(\omega)$ is the expected value with respect to the path probability density $\mu_P(\omega)$:
\begin{eqnarray}
\label{PathEnsembleAvg}
\mathbb{E}_P[f \,|\, x_0 =  x]
&=& \int_{\Omega_{\tau,  x}} \, \mu_P(\omega) \, f(\omega) \, \mathrm{d}\omega \cr	
&=& \int_{\Gamma} \int_{\Gamma} ... \int_{\Gamma} \, \mu_P( x_1,  x_2, ... ,  x_n \, | \,  x_0 =  x)\,  f( x_1,  x_2, ... ,  x_n) \, \times \cr
& \times & \, \mathrm{d} x_1, \mathrm{d} x_2, ... , \mathrm{d} x_n \, .
\end{eqnarray}

Typically, it is not possible to estimate the integral in eq.~\ref{PathEnsembleAvg}, however the ergodicity of the overdamped Langevin dynamics permits to use the \emph{Birkhoff-Khinchin's theorem} \cite{Cornfeld1982}.
Given a set of $m$ paths $S_{\tau,  x} = \lbrace \omega_1, \omega_2, ... \omega_m \rbrace \subset \Omega_{\tau,  x}$
generated by a discretization of eq.~\ref{eq:sde}, the expected value of the function $f$ is equal to the algebraic average of the values of the function $f$ evaluated over all the paths:
\begin{eqnarray}
    \mathbb{E}_P[f \,|\, x_0 =  x]  = \lim_{m\rightarrow \infty} \frac{1}{m} \sum_{\omega_k \in S_{\tau, x}} f(\omega_k) \, , 
\end{eqnarray}
where every path contributes with equal weight to the path ensemble average.

The path ensemble average of an observable $f$ depends on the path space and the path probability used to integrate eq.~\ref{PathEnsembleAvg}.
Thus we can use the Girsanov theorem to calculate the path ensemble average with respect to a measure $\tilde{P}$, using the path probability density $\mu_P(\omega)$:
\begin{eqnarray}
\mathbb{E}_{\widetilde{P}}[f \,|\, x_0 =  x] 
	&=& \int_{\Omega_{\tau,  x}} \, \widetilde{\mu}_P(\omega) \, f(\omega) \, \mathrm{d}\omega \cr
	&=& \int_{\Omega_{\tau,  x}} \, M_{\tau,  x}(\omega)\, \mu_P(\omega) \, f(\omega) \, \mathrm{d}\omega  \cr
	&=&\,\lim_{m\rightarrow \infty} \frac{1}{m} \sum_{\omega_k \in S_{\tau, x}} M_{\tau,  x}(\omega_k) f(\omega_k) \, .
\label{eq:pathEnsembleAverageReweighted}	
\end{eqnarray}
Note that not the path ensemble average as such is reweighted, but the weight with which every individual path contributes to the path ensemble average is scaled by $M_{\tau,  x}(\omega)$.
%

%
%
\subsection{Path reweighting for time-correlation functions and MSMs}
Consider two observable functions $a$ and $b$ defined on the state space $\Gamma$, the time-correlation function for a lag-time $\tau$  is defined as
\begin{eqnarray}
	\mathrm{cor}(a,b;\tau) &=& 	\int_\Gamma\int_\Gamma 
						a(x) \mu_\pi(x) p(x, y;\tau) b(y) 
						\, \mathrm{d}x\, \mathrm{d}y \cr
	&=&	\int_\Gamma a(x) \left[\int_\Gamma 
						 p(x, y;\tau) b(y) 
						\, \mathrm{d}y\right]\, \mu_\pi(x) \mathrm{d}x 				
\label{eq:correlationFunction}						
\end{eqnarray}
where $\mu_\pi(x):\Gamma \rightarrow \mathbb{R}$ is the stationary distribution:
\begin{eqnarray}
\mu_\pi(x) = \frac{\exp\left(-\beta V(x)\right)}{Z}
\label{eq:statDistr}
\end{eqnarray}
with $\beta = \nicefrac{1}{k_B T}$ and $Z=\int_\Gamma \exp\left(-\beta V(x)\right) \, \mathrm{d} x$ is the partition function.

The inner integral in eq.~\ref{eq:correlationFunction} can be regarded as a path ensemble average:
\begin{eqnarray}
	&&	\mathbb{E}_P[\, b(x_n) \,|\, x_0 =  x] \cr
    &=& \int_{\Omega_{\tau,x}} \mu_P(\omega) \, b(x_n) \, \mathrm{d}\omega \\
	&=& \int_{\Gamma}
	\left[\int_{\Gamma} \int_{\Gamma} \dots  \int_{\Gamma} \mu_P( x_1, x_2,..., x_n \, | \,  x_0= x) \, 
		\mathrm{d} x_1 \, \mathrm{d} x_2 \, ... \, \mathrm{d} x_{n-1}\right]
		b( x_n) \, \mathrm{d}  x_n \\
	&=& \int_{\Gamma} p( x,  y; \tau) \, b( y) \, \mathrm{d} y 	\, \, ,
\end{eqnarray}
while the outer integral in eq.~\ref{eq:correlationFunction} is a state space ensemble average:
\begin{equation}
\mathbb{E}_{\pi}[\, a(x)\mathbb{E}_P[\, b(x_n) \,|\, x_0 =  x] \,]  = \int_{\Gamma} a(x) \mathbb{E}_P[\, b(x_n) \,|\, x_0 =  x]
\mu_{\pi}(x) \, \mathrm{d}x \, .
\end{equation}

Thus the time-correlation function can be rewritten as 
\begin{eqnarray}
	\mathrm{cor}(a,b;\tau) = \mathbb{E}_{\pi}[a( x) \cdot \mathbb{E}_P[\,b(x_n) \,|\, x_0 =  x] ]\, .
\label{eq:correlationFunction2} 	
\end{eqnarray}

If we consider the perturbed potential $\tilde{V}(x) = V(x) + U(x)$, the associated path probability density $\mu_{\tilde{P}}$ and the stationary probability density $\mu_{\tilde\pi}$, the time-correlation function between $a(x)$ and $b(x)$ reads:
\begin{eqnarray}
	\tilde{\mathrm{cor}}(a,b;\tau) = \mathbb{E}_{\tilde\pi}[a( x) \cdot \mathbb{E}_{\tilde{P}}[\,b(x_n) \,|\, x_0 =  x] ]\, .
\label{eq:correlationFunction3} 	
\end{eqnarray}

Introducing two reweighting factors, one for the state space ensemble average and one for the path ensemble average, eq.~\ref{eq:correlationFunction3} can be written in terms of $\mu_{\pi}$ and $\mu_P$.
The first reweighting factor is the ratio between stationary probability densities:
\begin{equation}
g(x) = \frac{\mu_{\tilde\pi}( x)}{\mu_{\pi} ( x)} = \frac{Z}{\widetilde{Z}}\exp\left(-\beta U(x) \right)\, ,
\label{eq:RadonNikodym01}
\end{equation}
while the second one is the ratio between the path probability densities $M_{\tau,  x}(\omega)$, which can be explicitly calculated by eq.~\ref{eq:girsanov}:
\begin{eqnarray}
	\tilde{\mathrm{cor}}(a,b;\tau)&=&  \mathbb{E}_{{\pi}}[g( x) \cdot a( x) \cdot \mathbb{E}_{{P}}[\, M_{\tau,  x}(\omega) \cdot b(x_n) \,|\, x_0 =  x] ] \\
    &=& \int_{\Gamma} g( x) \mu_{\pi} ( x)  \, a( x) 
	 	\int_{\Omega_{\tau,  x}} M_{\tau,  x}(\omega) \mu_P(\omega) b( x_n) \, \mathrm{d}\omega
		\,\mathrm{d} x     \, .
\label{eq:correlationFunction4} 	
\end{eqnarray}

This result can be applied in MSMs, where the transition probability matrix of a stochastic process is constructed from a time-correlation matrix $\mathbf{C}(\tau)$.
Given a discretization of the state space $\Gamma$ in $N$ disjoint subsets $A_1,...,A_N$, the entries $C_{ij}$ of the matrix $\mathbf{C}(\tau)$ are time-correlation functions where $a(x)$ and $b(x)$ are replaced by the indicator functions  $a(x) = \mathbf{1}_{A_i}( x)$ and $b(x_n) = \mathbf{1}_{A_j}( x_n)$:
\begin{eqnarray}
	\tilde C_{ij}(\tau) 	
	&=& \int_{\Gamma} \mu_{\tilde\pi} ( x)  \, \mathbf{1}_{A_i}( x) 
	 	\int_{\Omega_{\tau,  x}}\mu_{\tilde P}(\omega) \mathbf{1}_{A_j}( x_n) \, \mathrm{d}\omega
		\,\mathrm{d} x  \cr
	&=& \int_{\Gamma} g( x) \mu_{\pi} ( x)  \, \mathbf{1}_{A_i}( x) 
	 	\int_{\Omega_{\tau,  x}} M_{\tau,  x}(\omega) \mu_{P}(\omega) \mathbf{1}_{A_j}( x_n) \, \mathrm{d}\omega
		\,\mathrm{d} x  \, .
\label{eq:correlationReweighted}		
\end{eqnarray}
Given a set of $m$ short paths $\lbrace \omega_1,...,\omega_m \rbrace \in S_{\tau, x}$, generated by integrating eq~\ref{eq:sde} with potential $V(x)$, $\tilde C_{ij}(\tau)$ is estimated as
\begin{eqnarray}
	\tilde C_{ij}(\tau) 	
	&=&	\lim_{m\rightarrow \infty }\frac{1}{m}\sum_{\omega_k \in  S_{\tau, x}}  
	    g([ x_0]_k) \mathbf{1}_{A_i}([ x_0]_k) \cdot M_{ x, \tau}(\omega_k) \mathbf{1}_{A_j}([ x_n]_k) \, .
	    \label{eq:corRewighted}
\end{eqnarray}
Finally the entries of the transition probability matrix between the subset $A_i$ and the subset $A_j$ for the perturbed dynamics are
\begin{eqnarray}
	\tilde T_{ij}(\tau) = \frac{\tilde C_{ij}(\tau)}{\sum_{j} \tilde C_{ij}(\tau)} \, . 
\label{eq:illustration09}	
\end{eqnarray}
Note that by normalizing the transition matrix the ratio of the partition functions $Z/\widetilde Z$, which in appears in eq.~\ref{eq:RadonNikodym01}, cancels and hence does not need to be calculated.
\section{The Square Root approximation of the infinitesimal generator}
\label{sec:sqra}
\subsection{The infinitesimal generator}
The time-evolution of the probability density $\rho(x,t)$ associated to the overdamped Langevin dynamics (eq.~\ref{eq:sde}) is given by the Fokker-Planck equation
\begin{eqnarray}
\partial_t \rho(x,t)  
&=& D \Delta \rho(x,t) + \nabla \left(\rho(x,t) \cdot \xi^{-1}  \mathbf{M}^{-1} \nabla V(x) \right)\cr
&=& \mathcal{Q}\rho(x,t)\,, 
\label{eq:FP}
\end{eqnarray}
where the operator $\mathcal{Q}$ is the \emph{infinitesimal generator} \cite{Lasota1994, Schuette1999b, Oksendal2003}.
Eq.~\ref{eq:FP} is also called the Smoluchowski diffusion equation.

Given an arbitrary discretization of the space in $N$ disjoint subsets, the entries of the \emph{rate matrix} $\mathbf{Q}$, i.e. the Galerkin discretization of $\mathcal{Q}$, reads
\begin{equation}
Q_{ij} = \frac{\langle \mathbf{1}_i \, ,\,  \mathcal{Q} \mathbf{1}_j \rangle_{\pi}}{\langle \mathbf{1}_i \, ,\,   \mathbf{1}_i \rangle_{\pi}} \, ,
\end{equation}
where $\mathbf{1}_i$ is the indicator function of the $i$th subset and the angle brackets denotes the weighted scalar product: $\langle u \vert v\rangle_{\pi} = \int u\, v\, \mu_\pi \mathrm{d}x$.
The resulting matrix representation of $\mathcal{Q}$ fulfills the master equation for a jump process:
\begin{equation}\label{eq:discr-time-FP}
\left.\frac{\partial p_i}{\partial \tau} \right\vert_{\tau=0} = C \sum_{i \sim j} (p_j Q_{ji} - p_i Q_{ij}) \, ,
\end{equation}
where $p_i, p_j$  are respectively the $i$th and $j$th elements of the vector $\mathbf{p}$, approximation of the probability density function $\rho$, $C$ is a normalization constant and the notation $i \sim j$ denotes neighboring subsets. 
The matrix $\mathbf{Q}$ has the properties of a rate matrix:
\begin{enumerate}
\item $Q_{ij}$ describes the transition rate from the set $j$ to the neighbor set $i$.
\item The diagonal elements satisfy $Q_{ii} = - \sum_{j \neq i} Q_{ij}$ and consequently the row-sums are zero $\sum_j Q_{ij}=0$.

\end{enumerate}

\subsection{Square root approximation}
Consider a Voronoi tessellation of the position space $\Gamma = \cup_{i=1}^{N} A_i$ in $N$ subsets and the transition rate matrix $\mathbf{Q}$ which is a discretization of the infinitesimal generator $\mathcal{Q}$.
The Gauss theorem allows to write the rate between adjacent subsets as \cite{Lie2013, Donati2018b}
\begin{equation}
\label{eq:gauss1}
Q_{ij,\,\mathrm{adjacent}} = \frac{1}{\pi_i} \oint_{\partial A_i \partial A_j} \Phi(z) \,  \mu_{\pi}(z) \, \mathrm{d} S(z) \, , 
\end{equation}
where $\pi_{i}=\int_{A_i}\mu_{\pi}(x) \, \mathrm{d} x$ is the probability that the system assumes a position $x\in A_i$, $\partial A_i \partial A_j$ is the intersecting surface between neighboring subsets $A_i$ and $A_j$ and $\Phi(z)$ denotes the flux of the configurations $z \in \partial A_i \partial A_j$, through the infinitesimal surface $\partial A_i \partial A_j$.

To approximate the surface integral in eq.~\ref{eq:gauss1}, we introduce three assumptions:
\begin{enumerate}
\item The flux does not depend on the position: $\Phi(x) = \Phi$.
\item The Voronoi subsets are so small that the potential energy $V(x)$ is almost constant within a subset: $V(x) |_{A_i} \approx V_i$.
This assumption applies also to the probability density within a subset $A_i$, then
\begin{eqnarray}
    \pi_i &=& \int_{A_i} \mu_\pi(x)\, \mathrm{d}x \,\approx\, \pi(x_i) \mathcal{V}_i \, ,
\label{eq:discretizedDensity}    
\end{eqnarray}
where $\mathcal{V}_i = \int_{A_i} 1 \,\mathrm{d}x$ is the volume of the subset $A_i$.
\item The potential and the probability density on the intersecting surface $\partial A_i \partial A_j$ are approximated respectively by the arithmetic average $ V(x) |_{\partial A_i \partial A_j} \approx \frac{V_i + V_j}{2}$, and the geometric average
\begin{eqnarray}
  \pi(z) = \frac{1}{Z}\exp \left(-\frac{1}{k_BT} \frac{V_i + V_j}{2} \right)
    &=& \sqrt{\pi(x_i) \pi(x_j)} \, .
\end{eqnarray}
\end{enumerate}

With these approximations, the integral in eq.\ref{eq:gauss1} becomes:
\begin{eqnarray}
\label{eq:gauss2}
Q_{ij\,\mathrm{adjacent}} & = &\frac{\Phi}{\pi(x_i) \mathcal{V}_i} \oint_{\partial A_i \partial A_j}  \,  \sqrt{\pi(x_i) \pi(x_j)} \, \mathrm{d} S(z) \cr
& = &\frac{\Phi}{\pi(x_i) \mathcal{V}_i} \mathcal{S}_{ij} \sqrt{\pi(x_i) \pi(x_j)}  \cr
& = & \Phi \,\frac{\mathcal{S}_{ij}}{\mathcal{V}_i}\sqrt{\frac{\pi(x_j)}{\pi(x_i)}}  \, ,
\, .
\end{eqnarray}
where $\mathcal{S}_{ij}$ is the measure of the intersecting surface.
Finally the entries of the rate matrix $\mathbf{Q}$ are written as
\begin{eqnarray}
Q_{ij} &=& 
\begin{cases}
\Phi \,\frac{\mathcal{S}_{ij}}{\mathcal{V}_i}\sqrt{\frac{\pi(x_j)}{\pi(x_i)}}   
                                            &\mbox{if  $i\ne j$, and  $A_i$ is adjacent to $A_j$}  \\
0                                           &\mbox{if  $i\ne j$, and  $A_i$ is not adjacent to $A_j$} \\
-\sum_{j=1, j\ne i}^n Q_{ij}                            &\mbox{if } i=j \, .
\end{cases} 
\label{eq:rate_matrix_01}
\end{eqnarray}
\subsection{Derivation of the flux}
The term $\Phi$ is derived from the Fick's second law \cite{Donati2021}, i.e. the Fokker-Planck equation for the overdamped Langevin dynamics with constant potential:
\begin{eqnarray}
\partial_t  \rho(x,t) &=&  D\Delta  \rho(x,t) = \mathcal{Q} \rho(x,t) \, .
\label{eq:FP_constV}    
\end{eqnarray}
Applying the Gauss theorem, the Laplacian of the probability density $\rho (x,t)$ over the small region $A_i$, is written as a surface integral of the gradient of the probability density
\cite{Arfken2001}:
\begin{equation}
    \Delta \rho (x_i,t) = \lim_{\mathcal{V_i} \rightarrow 0} \frac{1}{\mathcal{V}_i} \oint_{\mathcal{S}_i} \nabla \rho(z,t) \cdot \mathbf{n} \, \mathrm{d}S(z) \, ,
\label{eq:gauss3}
\end{equation}
where $\mathcal{V}_i$ and $\mathcal{S}_i$ are the volume and the surface of the subset $A_i$.

Approximating the gradient by the finite difference 
\begin{equation}
    \left.\nabla \rho(x,t) \right\vert_{x = x_j} \cdot \mathbf{n}_{ji}   \approx \frac{\rho(x_i, t) - \rho(x_j, t)}{h_{ji}} \, ,
\label{eq:rho_finite_difference}    
\end{equation}
and calculating the surface integral in eq.~\ref{eq:gauss3} , one obtains the master equation
\begin{eqnarray}
\partial_t  \rho_j(t)
&=& \sum_{i\sim j} D \frac{1}{h_{ij}}\frac{\mathcal{S}_{ij}}{\mathcal{V}_i}\rho_i(t) - D \frac{1}{h_{ij}}\frac{\mathcal{S}_{ij}}{\mathcal{V}_j}\rho_j(t) \cr
&=&\sum_{i\sim j} Q_{ij} \rho_i(t) - Q_{ji} \rho_j(t)
\label{eq:discFPE}
\end{eqnarray}
Because we assumed a constant potential $\sqrt{\frac{\pi(x_j)}{\pi(x_i)}} =1$, and comparing eq.~\ref{eq:gauss2} with eq.~\ref{eq:discFPE}, the flux is written as
\begin{equation}
    \Phi = \frac{D}{h_{ij}} \,,
\label{eq:flux02}    
\end{equation}{}
and the entries of the rate matrix can be further specified as
\begin{eqnarray}
Q_{ij} &=& 
\begin{cases}
\frac{D}{h_{ij}} \,\frac{\mathcal{S}_{ij}}{\mathcal{V}_i}\sqrt{\frac{\pi(x_j)}{\pi(x_i)}}   
                                            &\mbox{if  $i\ne j$, and  $A_i$ is adjacent to $A_j$}  \\
0                                           &\mbox{if  $i\ne j$, and  $A_i$ is not adjacent to $A_j$} \\
-\sum_{j=1, j\ne i}^n Q_{ij}                            &\mbox{if } i=j \, .
\label{eq:finalRateMatrix}
\end{cases} 
\end{eqnarray}
For infinitely small subsets the rate matrix so defined converges to the Fokker-Planck operator as was proven in refs.~\cite{Heida2018, Donati2018b, Donati2021}.
\subsection{Dynamical reweighting by SqRA}
Consider a system governed by potential energy function $V(x)$ and a perturbed potential $\tilde V(x)$ as defined in eq.~\ref{eq:perturbedPotential}.
According to eq.~\ref{eq:finalRateMatrix}, the entries of the perturbed rate matrix $\tilde{\mathbf{Q}}$ between adjacent subsets read
\begin{eqnarray}
\tilde{Q}_{ij\,\mathrm{adjacent}}
& = & \Phi \,\frac{\mathcal{S}_{ij}}{\mathcal{V}_i}\,\sqrt{\frac{\tilde{\pi}(x_j)}{\tilde{\pi}(x_i)}} \cr
& = & \Phi \,\frac{\mathcal{S}_{ij}}{\mathcal{V}_i}\,\sqrt{\frac{\exp[-\beta V(x_j) - \beta U(x_j)]}{\exp[-\beta V(x_i) - \beta U(x_i)]}} \, .
\end{eqnarray}
Factorizing the exponential terms, one obtains 
\begin{eqnarray}
\tilde{Q}_{ij\,\mathrm{adjacent}}
& = & \Phi \,\frac{\mathcal{S}_{ij}}{\mathcal{V}_i} \, 
\sqrt{\frac{\pi(x_j)}{\pi(x_i)}}
\,\sqrt{\frac{\exp(- \beta U(x_j))}{\exp(- \beta U(x_i))}} \cr
& = & Q_{ij\,\mathrm{adjacent}}\,\sqrt{\frac{\exp(- \beta U(x_j))}{\exp(- \beta U(x_i))}} \,,
\label{eq:sqraReweighting}
\end{eqnarray}
where $\mathbf{Q}$ is the transition rate matrix of the unperturbed system.
Then, eq.~\ref{eq:sqraReweighting} defines a reweighting scheme for transition rates.

\subsection{The relationship between SqRA and MSM}
The MSM transition probability matrix $\mathbf{T}(\tau)$ defined in eq.~\ref{eq:illustration09} is the Galerkin discretization of the transfer operator $\mathcal{T}(\tau)$ which propagates weighted probability densities $u_t(x)= \rho(x,t) / \mu_\pi(x)$, where  $\rho(x,t)$ is the probability density at time $t$ introduced in eq.~\ref{eq:FP}, and $\mu_\pi(x)$ is the stationary probability density defined in eq.~\ref{eq:statDistr}.
The operator $\mathcal{T}(\tau)$ is related to the infinitesimal generator $\mathcal{Q}$, defined in eq.~\ref{eq:FP}, by the relationship
\begin{eqnarray}
\mathcal{Q}
& = & \left. \frac{\partial \mathcal{T}(\tau)}{\partial \tau} \right\vert_{\tau=0}\cr
& = & \lim_{\tau \downarrow 0} \frac{\mathcal{T}(\tau) - \mathcal{T}(0)}{\tau} \, ,
\label{eq:der1}
\end{eqnarray}

However, the analogous relationship
\begin{eqnarray}
\mathbf{Q}^{\mathrm{SqRA}} \, 
&=&
\lim_{\tau \downarrow 0} \frac{
\mathbf{T}^{\mathrm{MSM}}(\tau)
-
\mathbf{T}^{\mathrm{MSM}}(0) 
}{\tau}
\, ,
\label{eq:der2}
\end{eqnarray}
between the corresponding transition matrices  respectively built by MSM and SqRA, is more delicate and not always accurate.

The MSM construction requires indeed discretizing the space into subsets, which results in the loss of the Markovian property \cite{Prinz2011}.
This is a systematic error, i.e. the dynamics represented by the MSM could significantly differ from the true dynamics.
The problem can be mitigated in two ways: (i)
by a fine discretization of the space; (ii) by choosing a large enough lag time $\tau$ that satisfies the Chapman-Kolmogorov equation $\mathbf{T}(n \cdot \tau) = \mathbf{T}(\tau)^n$.

The first option increases the statistical error due to poor sampling of each subset.
Similarly, the same problem affects the SqRA construction if the stationary distribution in eq.~\ref{eq:finalRateMatrix} is approximated by a histogram.

In the second case, the definition of  infinitesimal generator is violated, since it requires taking the limit $\tau \downarrow 0$.
The transition rate matrix, as defined in  eq.~\ref{eq:der2}, expresses instantaneous transition probabilities between adjacent subsets, and then requires a short lag time $\tau$.

In conclusion, while it is correct the relationship between operators in eq.~\ref{eq:der1}, the relationship defined in eq.~\ref{eq:der2} between the corresponding matrices is not well defined.
For a more detailed discussion about the relationship between MSM and SQRA we refer to Ref.~\cite{Donati2018b}.

\section{Numerical experiments}
\subsection{Transformation of the  M\"uller-Brown potential}
\label{sec:resA}
As first example of application, we studied a convex combination of potentials based on the $\lambda$-dependent potential
\begin{eqnarray}
V(x,y; \, \lambda) = \lambda V_A(x,y) + (1-\lambda) V_B(x,y) \, ,
\label{eq:lambda_potential}
\end{eqnarray}
where both $V_A(x,y)$ and $V_B(x,y)$ are defined by the same function
\begin{eqnarray}
\sum_{n=1}^2 
A_n \exp\left(a_n (x - x_n)^2 + b_n(x - x_n)(y - y_n) + c_n(y - y_n)^2\right) \, ,
\end{eqnarray}
but using two different sets of parameters reported in table \ref{tab:tab1}.
\begin{table}[!ht]
\begin{tabular}{|ccccc|}
\hline
\multicolumn{1}{|c|}{}
 & \multicolumn{2}{c|}{$V_A(x,y)$} & \multicolumn{2}{c|}{$V_B(x,y)$} \\ \hline
\multicolumn{1}{|c|}{} & $n=1$ &  \multicolumn{1}{|c|}{$n=2$} & $n=1$ & \multicolumn{1}{|c|}{$n=2$} \\ \hline\hline
\multicolumn{1}{|c|}{$A_n$} & -20 & \multicolumn{1}{c|}{-10} & -17 & \multicolumn{1}{c|}{1.5} \\ 
\multicolumn{1}{|c|}{$a_n$} & -1  & \multicolumn{1}{c|}{-1} & -6.5 & \multicolumn{1}{c|}{0.7} \\ 
\multicolumn{1}{|c|}{$b_n$} & 0 & \multicolumn{1}{c|}{0} & 11 & \multicolumn{1}{c|}{0.6} \\ 
\multicolumn{1}{|c|}{$c_n$} & -10 & \multicolumn{1}{c|}{-10} & -6.5 & \multicolumn{1}{c|}{0.7} \\ 
\multicolumn{1}{|c|}{$x_n$} & 1 & \multicolumn{1}{c|}{0} & -0.5 & \multicolumn{1}{c|}{-1} \\ 
\multicolumn{1}{|c|}{$y_n$} & 1 & \multicolumn{1}{c|}{0.5} & 1.5 & \multicolumn{1}{c|}{1} \\ 
\hline\hline
\end{tabular}
\caption{Parameters of the potentials $V_A(x,y)$ and $V_B(x,y)$.}
\label{tab:tab1}
\end{table}

\begin{figure*}[ht]
\includegraphics[]{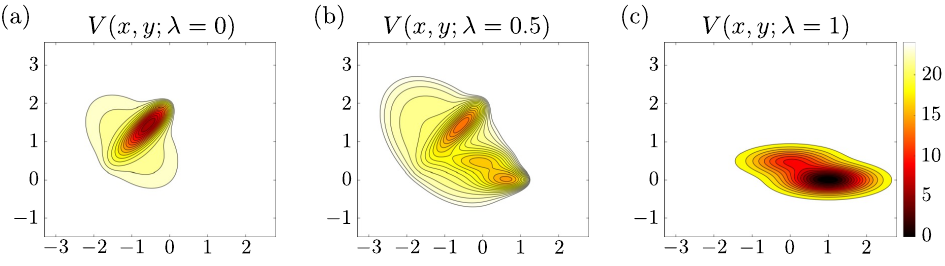}
\caption{Potential energy function: $\lambda = 0$ (a), $\lambda = 0.5$ (b) and $\lambda = 1$ (c).}
\label{fig:fig2}
\end{figure*}

The two potentials are illustrated in fig.~\ref{fig:fig2}-a,c: the initial state ($\lambda = 0$) is a basin with  minimum located at $(-1,1.5)$; the final state ($\lambda = 1$) is composed by a deep basin around $(1,-0.2)$ and a shallow minimum at $(0,0.5)$.
The intermediate state ($\lambda = 0.5$ in fig.~\ref{fig:fig2}-b) is the M\"uller-Brown potential \cite{Muller1979}, characterized by the global and local minima of the states $A$ and $B$.

In our numerical experiment, we generated a set of trajectories at potential $V(x,y;\,\lambda^{\mathrm{sim}}=0.5)$ and we built the MSMs at different $\lambda^{\mathrm{target}}$ values in $[0,1]$ via Girsanov reweighting.
This choice is motivated by the fact that the regions kinetically more relevant, i.e. with low potential energy, of the potentials $V(x,y;\,\lambda=0)$ and $V(x,y;\,\lambda=1)$ do not overlap.
Consequently, the associated probability densities barely satisfy the condition of absolute continuity (eq.~\ref{eq:absContPath}) and a direct reweighing $V(x,y;\,\lambda=0) \rightarrow V(x,y;\,\lambda=1)$, either vice-versa, would produce low quality results.
On the contrary, the potential $V(x,y;\,\lambda=0.5)$ encompasses both the potentials and allows for reweighting in both directions: $V(x,y;\,\lambda=0.5) \rightarrow V(x,y;\,\lambda=0)$ and $V(x,y;\,\lambda=0.5) \rightarrow V(x,y;\,\lambda=1)$.
To verify the results obtained by Girsanov reweighting, we built SqRA rate matrices as reference solutions.

\subparagraph{Methods}
We first used the Euler-Maruyama scheme \cite{Leimkuhler2015} with an integrator timestep $\Delta t = 5\times 10^{-4} \, \mathrm{ps}$ to solve eq.~\ref{eq:sde} at potential $V(x,y; \, \lambda^{\mathrm{sim}} = 0.5)$, generating five long trajectories of length $1\times10^7$ timesteps.
The potential has units of kJ/mol.
The mass and the friction in eq.~\ref{eq:sde} were respectively $m = 1\, \mathrm{amu}$ and $\xi = 1\, \mathrm{ps}^{-1}$.
The temperature of the system was $T=300\, \mathrm{K}$ and the thermodynamic beta was $\beta = \nicefrac{1}{k_B T}= 0.40 \, \mathrm{kJ^{-1}}$ with the molar  Boltzmann constant $k_B = 0.008314463 \,\mathrm{kJ\cdot mol^{-1}\cdot K^{-1}}$.
%

%

The state space was discretized in $K=300$ Voronoi subsets applying the $K$-means clustering algorithm to one of the trajectories, but enforcing a uniform selection of the centers to guarantee Voronoi subsets of approximately the same size.
The same tessellation of the space was used to build both the MSM transition probability matrices and SqRA rate matrices.

To construct the MSMs, we counted the transitions between Voronoi subsets within a certain lag time $\tau$ selected in a range between 0 and 0.35 ps.
Detailed balance was enforced by symmetrizing the resulting $300\times300$-count matrix: 
$\mathbf{C}_{\mathrm{sym}}(\tau) = \mathbf{C}(\tau) + \mathbf{C}^{\top}(\tau)$.
The MSM transition probability matrix $\mathbf{T}(\tau)$ was obtained by row-normalizing $\mathbf{C}_{\mathrm{sym}}(\tau)$. 
The MSM procedure was repeated for each trajectory, then we estimated the eigenvectors and the implied timescales as averages accompanied by the standard deviation.

The procedure to build the MSM for any $\lambda$ value by dynamical reweighting was the same, but each transition was weighted by the product of the terms defined in eqs.~\ref{eq:girsanov} and \ref{eq:RadonNikodym01}, where the potential $U$ is the difference between target potential and simulated potential.
For this purpose, we estimated on-the-fly and stored the following terms at each timestep of the simulations:
\begin{enumerate}
    \item The gradients $\nabla V_A(x,y)$ and $\nabla V_B(x,y)$,
    \item The energy $V_A(x,y)$ and $V_B(x,y)$,
    \item The random numbers generated to approximate the stochastic term in eq.~\ref{eq:sde}.
\end{enumerate}
Afterward, we calculated the energy difference $U(x,y; \, \lambda^{\mathrm{target}})$ and its gradient respectively as
\begin{eqnarray}
U(x,y; \, \lambda^{\mathrm{target}}) 
&=&
V(x,y; \,  \lambda^{\mathrm{target}}) - V(x,y; \,  \lambda^{\mathrm{sim}}) \cr
&=& 
\left(\lambda^{\mathrm{target}} -  \lambda^{\mathrm{sim}}\right) V_A(x,y)  +\cr
&&+ \left( \lambda^{\mathrm{sim}} - \lambda^{\mathrm{target}} \right)   V_B(x,y) \, ,
\end{eqnarray}
and
\begin{eqnarray}
\nabla U(x,y; \, \lambda^{\mathrm{target}}) 
&=&
\nabla V(x,y; \,  \lambda^{\mathrm{target}}) - \nabla V(x,y; \,  \lambda^{\mathrm{sim}}) \cr
&=& 
\left(\lambda^{\mathrm{target}} -  \lambda^{\mathrm{sim}} \right) \nabla V_A(x,y)  +\cr 
&&+ \left( \lambda^{\mathrm{sim}} - \lambda^{\mathrm{target}} \right)   \nabla V_B(x,y) \, .
\end{eqnarray}
Eigenvectors and eigenvalues of the MSM transition matrix $\mathbf{T}(\tau)$ and the SqRA rate matrix $\mathbf{Q}$ were obtained using the eigenvalue solver implemented in MATLAB.

\begin{figure*}[ht]
\includegraphics[]{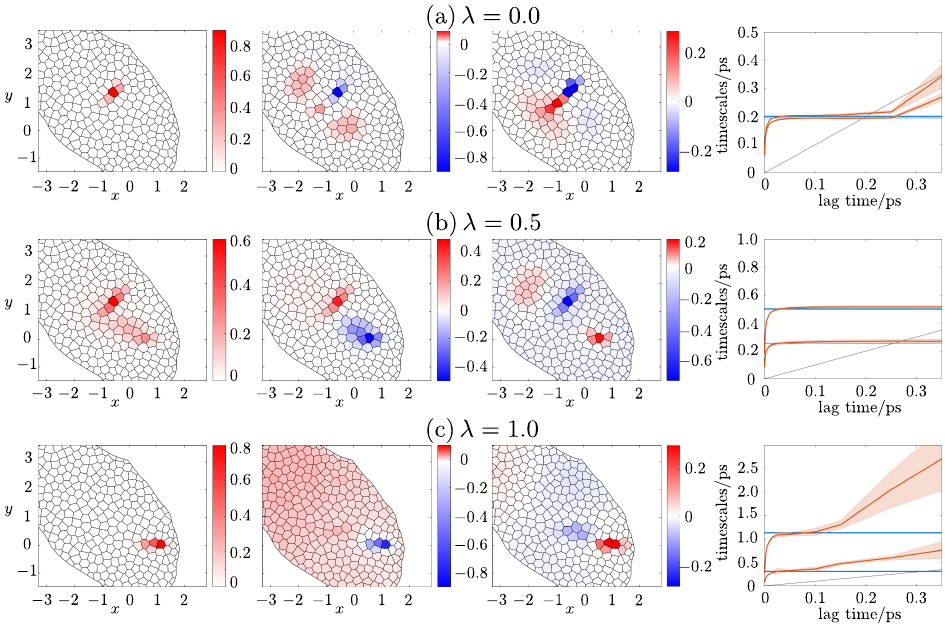}
\caption{
First three left MSM eigenvectors and first two MSM/SqRA implied timescales as functions of the lag time.
Each row corresponds to a two-dimensional system determined respectively by $\lambda=0.0$ (a), $\lambda = 0.5$ (b) and $\lambda = 1.0$ (c) in eq.~\ref{eq:lambda_potential}.
The timescales figures include both the MSM implied timescales (red color with shaded error bars) and the SqRA implied timescales (blue).
}
\label{fig:fig3}
\end{figure*}

\subparagraph{Eigenvectors and implied timescales}
The first three left MSM eigenvectors of the simulated potential with $\lambda^{\mathrm{sim}}=0.5$ are plotted in fig.~\ref{fig:fig3}-b.
We present the left eigenvectors of $\mathbf{T}(\tau)$ and $\mathbf{Q}$, where both matrices are defined to be row-normalized.
The first eigenvector represents the stationary distribution, while the second and third left eigenvectors are the slowest kinetic modes that contribute to the dynamics of the system.
The associated implied timescales, respectively of 0.50 ps and 0.25 ps, quickly converge at $\tau = 0.01$ ps, indicating that the discretization error is negligible.
The SqRA generated the same left eigenvectors and implied timescales (blue solid lines) are in excellent agreement with the MSMs prediction.

In fig.~\ref{fig:fig3}-a,c we report the left eigenvectors and the implied timescales for the target potentials with $\lambda^{\mathrm{target}}=0$ and $\lambda^{\mathrm{target}}=1$, respectively obtained reweighted the trajectories generated at potential with $\lambda^{\mathrm{sim}} = 0.5$.
As the dynamics is fully located respectively within the basins $V_A(x,y)$ and $V_B(x,y)$, the first left eigenvectors show single peaks associated with the minima of the potential states.
For $\lambda^{\mathrm{target}} = 0$ the second and third eigenvectors describe two kinetic modes occurring within the basin and associated to the same timescale of 0.19 ps.
For $\lambda^{\mathrm{target}} = 1$ the slowest revealed process is between the basin $V_B(x,y)$ and the outer region, while the second slowest process is between the global and the local minimum of the potential; the associated timescales are respectively 1.15 ps and 0.36 ps.

The SqRA matrices of the two target potentials yielded the eigenvectors that are virtually indistinguishable from the MSM eigenvectors and are therefore not shown. 
The implied timescales calculated from the SqRA rate matrix are in good agreement with the MSM implied timescales (blue solid lines). 
However, our results show that the path reweighting procedure is sensitive to the choice of the MSM lag time.
For $\lambda^{\mathrm{target}} = 0$, the standard deviation is small until a lag time of 0.25 ps, while for $\lambda^{\mathrm{target}} = 1$ the implied timescales and their standard deviation diverge at a lag time of 0.1 ps.
At large lag times $\tau$, the reweighting factor in the path integral in in eq.~\ref{eq:girsanov} can become numerically unstable.
In the second case, the instability is also augmented by the difference potential $U(\lambda^{\mathrm{target}} = 1)$ which is greater than $U(\lambda^{\mathrm{target}} = 0)$: the difference between the minimum of $V(\lambda=0)$ and the corresponding minimum of $V(\lambda=0)$ is approximately $- 8 \, \mathrm{kJ}$; while the difference between the minimum of $V(\lambda=1)$ and the corresponding minimum of $V(\lambda=0)$ is approximately $- 13 \, \mathrm{kJ}$.
However, the implied timescales are stable within a lag time window of 0.1 ps, making the reweighting results still acceptable.

\begin{figure*}[ht]
\includegraphics[]{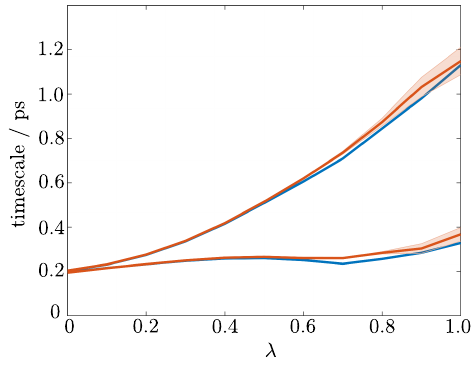}
\caption{First two MSM (red) and SqRA (blue) implied timescales as functions of $\lambda$.}
\label{fig:fig4}
\end{figure*}

Following the same procedure we were able to efficiently build the MSM also for intermediate $\lambda$ values.
The first two implied timescales as functions of the parameter $\lambda$ obtained at lag time $\tau=0.1$ ps are illustrated in fig.~\ref{fig:fig4}.
The leading implied timescales overlap at low $\lambda$, but they split as the state $V_A$ induces the system to a metastability, characterized by a rare transition between the basins.
They are in excellent agreement with the SqRA implied timescales, nonetheless the standard deviation increases with $\lambda > 0.5$.

%
\subsection{Dynamical reweighting of force field parameters}
\label{sec:resB}
\begin{figure*}[ht]
\includegraphics[]{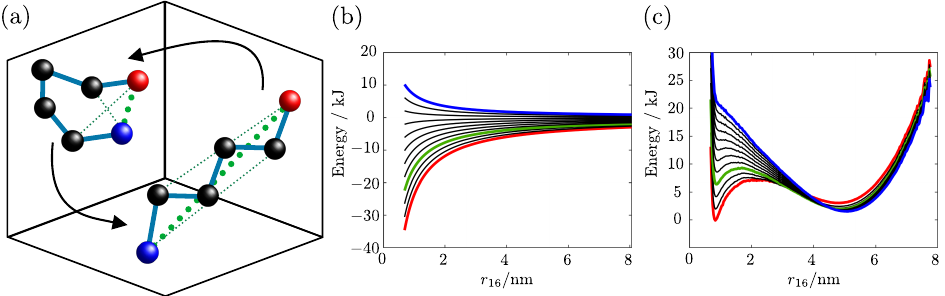}
\caption{6-atoms molecule. (a) Cartoon of the molecule:  close and open conformation; (b) Coulomb interaction and (c) free energy profile along the reaction coordinate for different electric constant values: $k_{ele} =$ -50 (blue), 0 (green), 150 (red) $\mathrm{kJ \cdot nm \cdot mol^{-1} \cdot q_e^{-2}}$.}
\label{fig:fig5}
\end{figure*}

As second example, we studied the prototype of a molecule of six atoms illustrated in fig.~\ref{fig:fig5}-a.
The solid lines represent the bond distances between atoms, while the dotted lines represent the interaction between non-bonded atoms.
The end atoms, colored in red and blue, have respectively an electric charge $q_1 = +0.39 \, q_e$ and $q_6 = -0.39 \, q_e$, where $q_e$ is the elementary charge; the charges of the atoms colored in black were all equal to zero.
Each atom has respectively a mass $m=12\, \mathrm{amu}$ and a friction constant $\xi = 10 \, \mathrm{ps^{-1}}$.
The force field of the molecule is defined by the potential energy function
\begin{eqnarray}
V
&=& V_{\mathrm{bonds}} + V_{\mathrm{angles}} + V_{\mathrm{torsion}} +  V_{\mathrm{nonbonded}}  \cr
&=&
\sum_{\substack{
  ij=\lbrace 1,2 \rbrace,
\lbrace 2,3 \rbrace,
 \\ 
\lbrace 3,4 \rbrace,
\lbrace 4,5 \rbrace,
\lbrace 5,6 \rbrace
}
}\frac{1}{2}k_{ij}\left(r_{ij} - r_{0,ij}\right)^2 + \cr
&&+
\sum_{\substack{
  ijk=\lbrace 1,2,3 \rbrace,
\lbrace 2,3,4 \rbrace,
 \\ 
\lbrace 3,4,5 \rbrace,
\lbrace 4,5,6 \rbrace
}}
\frac{1}{2}k_{ijk}\left(\theta_{ijk} - \theta_{0,ijk}\right)^2 + \cr
&&+\sum_{\substack{
  ijkl=\lbrace 1,2,3,4 \rbrace,
 \\ 
\lbrace 2,3,4,5 \rbrace,
\lbrace 3,4,5,6 \rbrace
}}
k_{ijkl}\cos\left(\Psi_{ijkl} - \Psi_{0,ijkl}\right) + \cr
&& 
+ 
\sum_{\substack{
  ij=\lbrace 1,5 \rbrace,
 \\ 
\lbrace 1,6 \rbrace,
\lbrace 2,6 \rbrace
}
} \varepsilon
\left[
\left(\frac{r_{0,ij}}{r_{ij}}\right)^{12} 
-2
\left(\frac{r_{0,ij}}{r_{ij}}\right)^{6} 
\right]+ \cr
&& 
+ k_{ele} \frac{q_1 q_6}{r_{16}} 
 \,,
\end{eqnarray}
where each term governs respectively the motion of the rigid bond lengths $r_{ij}$, the bond angles $\theta_{ijk}$, the torsions $\Psi_{ijkl}$, the Lennard-Jones and Coulomb potential between non bonded atoms.
The complete set of parameters is reported in tab.~\ref{tab:tab2} in Appendix \ref{sec:appendix2}.

The dynamics of the system is dominated by the stretching-compressing of the molecule along the Euclidean distance $r_{16}$ between the non-bonded atoms $1$ and $6$, which was chosen as reaction coordinate.
Here, we are interested in understanding the contribution of the Coulomb potential,  illustrated in fig.~\ref{fig:fig5}-b for different electric constant values, to this process.
%
For this purpose, we simulated the system at $k_{ele}^{\mathrm{sim}} = 100 \, \mathrm{kJ \cdot nm \cdot mol^{-1} \cdot q_e^{-2}}$, then we built the MSM for different $k_{ele}^{\mathrm{target}}$ values using both the Girsanov and the SqRA reweighting techniques.

\subparagraph{Methods}
We solved eq.~\ref{eq:sde} using the Euler-Maruyama scheme \cite{Leimkuhler2015} with an integrator timestep $\Delta t = 2\times 10^{-3} \, \mathrm{ps}$, generating five long trajectories of length $5\times10^8$ timesteps, saving the positions every $\mathsf{nstxout}= 10$ timesteps.
The Coulomb potential was set with an electric constant $k_{ele} = 100 \, \mathrm{kJ \cdot nm \cdot mol^{-1} \cdot q_e^{-2}}$.

To build the MSMs, we discretized the one-dimensional reaction coordinate $r_{16}$ in 100 small equal intervals and we built the matrix $\mathbf{C}(\tau)$ counting the transitions within a lag time $\tau$ chosen in a range between 0 and 50 ps, then we estimated the transition probability matrix $\mathbf{T}(\tau)$ dividing each entry of the matrix $\mathbf{C}(\tau)$ by the sum of the rows.
We enforced detailed balance symmetrizing the matrix $\mathbf{C}(\tau)$ as described in the previous example.

To perform the path reweghting, during the simulations we saved the terms
\begin{eqnarray}
U(r_{16}) =  \frac{q_1 q_6}{r_{16}} \, ,
\label{eq:Ur16}
\end{eqnarray}
every $\mathsf{nstxout}$ timesteps, i.e. at the same frequency of the atom positions.
In addition, every $\mathsf{nstxout}$ timesteps, we calculated and estimated on-the-fly the stochastic integral
\begin{eqnarray}
I_S &=&
\sum_{n = 1}
^{\mathsf{nstxout}}
\frac{\nabla_{\mathbf{r}_1} U(r_{16}(n))}
{m_1 \xi_1 \sigma_1}
\, \eta_{\mathbf{r}_1}(n) \sqrt{\Delta t} 
+\cr
&&+
\sum_{n = 1}
^{\mathsf{nstxout}}
\frac{\nabla_{\mathbf{r}_6} U(r_{16}(n))}
{m_6 \xi_6 \sigma_6}
\, \eta_{\mathbf{r}_6}(n) \sqrt{\Delta t}
\, ,
\label{eq:IS}
\end{eqnarray}
and the Riemann integral
\begin{eqnarray}
I_R &=&
\sum_{n = 1}
^{\mathsf{nstxout}}
\frac{
\nabla_{\mathbf{r}_1} U(r_{16}(n))
\cdot
\nabla_{\mathbf{r}_1} U(r_{16}(n))
}
{m_1^2 \xi_1^2 \sigma_1^2}
\,  \Delta t 
 +\cr
&&+
\sum_{n = 1}
^{\mathsf{nstxout}}
\frac{
\nabla_{\mathbf{r}_6} U(r_{16}(n))
\cdot
\nabla_{\mathbf{r}_6} U(r_{16}(n))
}
{m_6^2 \xi_6^2 \sigma_6^2}
\, \Delta t
\, ,
\label{eq:IR}
\end{eqnarray}
where $\nabla_{\mathbf{r}_1} U$ and $\nabla_{\mathbf{r}_6} U$ denote respectively the gradient of the potential $U(r_{16})$ with respect to the Cartesian coordinates of atom 1 and 6.
Correspondingly, $\eta_{\mathbf{r}_1}$ and $\eta_{\mathbf{r}_6}$ are two three-dimensional vectors containing the random numbers used to integrate the equations of the motion along the coordinates of atom 1 and 6.
Note that in eqs.~\ref{eq:Ur16}, \ref{eq:IS} and \ref{eq:IR}, we omitted the target electric constant $k_{ele}^{\mathrm{target}}$, and we scaled the reweighting terms during the post-analysis.
This permitted the application of the path-reweighting for several $k_{ele}$ values, however, attention must be paid to the choice of the scaling parameter: if the simulation was run with an electric constant $k_{ele}^{\mathrm{sim}}$ and one desires to reweight the simulation at the target parameter $k_{ele}^{\mathrm{target}}$, eqs.~\ref{eq:Ur16}, \ref{eq:IS} must be multiplied by the difference
\begin{eqnarray}
k_{ele}^{\mathrm{diff}} = k_{ele}^{\mathrm{target}} - k_{ele}^{\mathrm{sim}}\, ,
\label{eq:kelediff}
\end{eqnarray}
and eq.~\ref{eq:IR} must be multiplied by the square $k_{ele}^{\mathrm{diff}\,2}$.

\begin{figure*}[ht]
\includegraphics[]{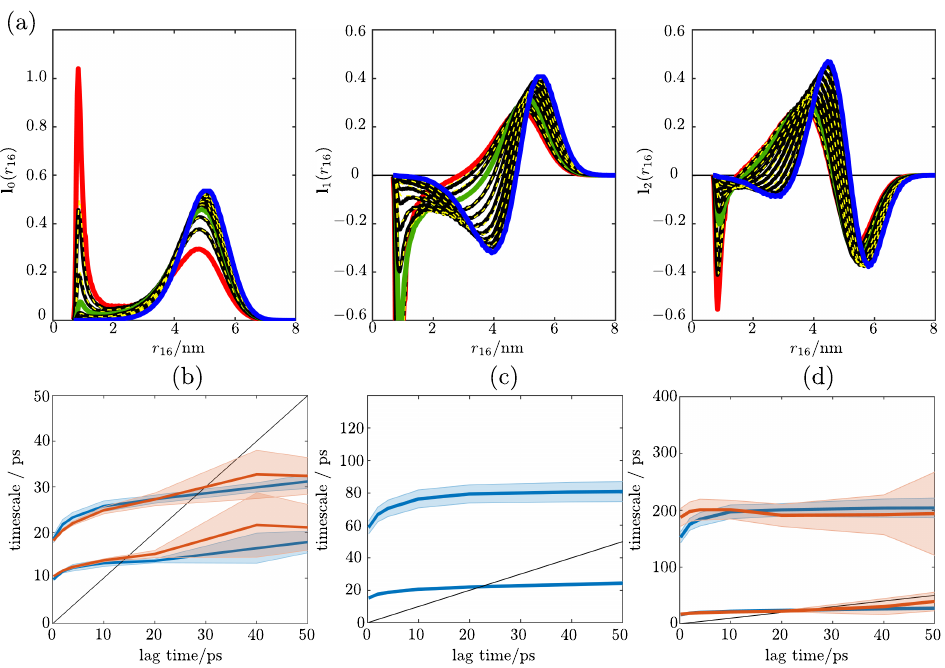}
\caption{6-atoms molecule. (a) First three MSM (black solid lines) and SqRA (yellow dashed lines) left eigenvectors $\mathbf{l}_0(r_{16})$, $\mathbf{l}_1(r_{16})$ and $\mathbf{l}_2(r_{16})$ for different values of the electric constant: $k_{ele} =$ -50 (blue), 0 (green), 150 (red) $\mathrm{kJ \cdot nm \cdot mol^{-1} \cdot q_e^{-2}}$. 
(b,c,d) First two MSM implied timescales at $k_{ele} =$ -50 (a), 0 (b), 150 (c) $\mathrm{kJ \cdot nm \cdot mol^{-1} \cdot q_e^{-2}}$ by reweighting (red) and direct simulation (blue).
}
\label{fig:fig6}
\end{figure*}

In this experiment, we also used the dynamical reweighting scheme based on SqRA, applying eq.~\ref{eq:sqraReweighting}.
First, we approximated the weighted stationary distribution along the reaction coordinate calculating the histogram of the simulation projected along the reaction coordinate $r_{16}$, then we multiplied it by
\begin{eqnarray}
\exp\left(-\beta k_{ele}^{\mathrm{diff}} U (r_{16})\right) \, ,
\end{eqnarray}
where $k_{ele}^{\mathrm{diff}}$ is defined in eq.~\ref{eq:kelediff} and $U (r_{16})$ is defined in eq.~\ref{eq:Ur16}.
In building the rate matrix (eq.~\ref{eq:finalRateMatrix}), we omitted the unknown diffusion constant $\hat{D}$ along the reaction coordinate which appears in the flux (eq.~\ref{eq:flux02}).
It follows that SqRA rate matrix provides the correct left eigenvectors $\mathbf{l}_i$ which satisfy the eigenvalue equation
\begin{eqnarray}
    \mathbf{l}_i^{\top}\mathbf{Q} &=& \tilde{\kappa}_i \mathbf{l}_i^{\top} \quad \mathrm{with}\ i\geq 0.
\end{eqnarray}
However the eigenvalues $\tilde{\kappa}_i$ have length units $[\mathrm{nm^{-2}}]$, and correspondingly the quantities
\begin{eqnarray}
\tilde{t}_i^{\, \mathrm{SqRA}} = -\frac{1}{\tilde{\kappa}_i}\, ,
\end{eqnarray}
have units $[\mathrm{nm^{2}}]$, and do not represent the correct SqRA implied timescales.
To solve this, we calculated the diffusion constant $\hat{D}$ along the reaction coordinate, with units $[\mathrm{nm^2 \cdot ps^{-1}}]$, from the ratio
\begin{eqnarray}
\hat{D} = \frac{\tilde{t}_1^{\, \mathrm{SqRA}}}{t_1^{\mathrm{MSM}}} \, ,
\label{eq:Diffu}
\end{eqnarray}
where $t_1^{\mathrm{MSM}}$ is the first MSM implied timescales with units $[\mathrm{ps}]$.
Then, the physically meaningful SqRA implied timescales is given by
\begin{eqnarray}
t_n^{\mathrm{SqRA}} = \hat{D} \cdot \tilde{t}_n^{\, \mathrm{SqRA}} \, .
\label{eq:tn}
\end{eqnarray}
Note that eq.~\ref{eq:tn} applies for any timescale $t_n$ as the ratio between the SqRA $\tilde{\kappa}_i$ is preserved \cite{Lie2013}.
%
\subparagraph{Eigenvectors}
The first three MSM eigenvectors $\mathbf{l}_0(r_{16})$, $\mathbf{l}_1(r_{16})$ and $\mathbf{l}_2(r_{16})$ of the simulated potential ($k_{ele} = 100 \, \mathrm{kJ \cdot nm \cdot mol^{-1} \cdot q_e^{-2}}$) are drawn with green color in fig.~\ref{fig:fig6}-a, and show that the system can assume both the close and open conformation, but with a propensity for the latter. 
The corresponding implied timescales, which are reported in fig.~\ref{fig:fig6}-c, converge at a lag time of 10 ps, implying a good state space discretization.
This simulation was then reweighted to different values of the electric constant in the range $k_{ele} \in [-50,150]\, \mathrm{kJ \cdot nm \cdot mol^{-1} \cdot q_e^{-2}}$, using both the Girsanov and SqRA techniques.
The corresponding left eigenvectors $\mathbf{l}_0(r_{16})$, $\mathbf{l}_1(r_{16})$ and $\mathbf{l}_2(r_{16})$ of the reweighted MSMs perfectly overlap with the reweighted SqRA left eigenvectors as shown in fig.~\ref{fig:fig6}-a (respectively black solid lines and yellow dashed lines).
The eigenvectors show that the system, for negative or weakly positive $k_{ele}$ values ($k_{ele}> 50\, \mathrm{kJ \cdot nm \cdot mol^{-1} \cdot q_e^{-2}}$), is stable in a open conformation with $r_{16} \approx 5 \, \mathrm{nm}$; on the contrary, at $k_{ele}< 50\, \mathrm{kJ \cdot nm \cdot mol^{-1} \cdot q_e^{-2}}$, the attractive force between the end atoms is not negligible and the system becomes metastable, i.e. it can assume both the open and close conformations.
These considerations can also be drawn from the free energy profile, derived by reversing the first left eigenvector along the reaction coordinate (fig.~\ref{fig:fig5}-c).

In order to validate our results, we verified the convergence of the reweighted MSMs and their agreement with test simulations at $k_{ele} = 0$ and $150\, \mathrm{kJ \cdot nm \cdot mol^{-1} \cdot q_e^{-2}}$.
The eigevectors from direct simulations are highlighted respectively with blue and red color, and perfectly overlap with the eigenvectors predicted by the Girsanov and SqRA reweighting.
The MSM implied timescales plotted in fig.~\ref{fig:fig6}-a,c (blue solid lines), are also in good agreement with the reweighted implied timescales (red solid lines) and both converge at a lag time of 10 ps.
%

\begin{figure*}[ht]
\includegraphics[]{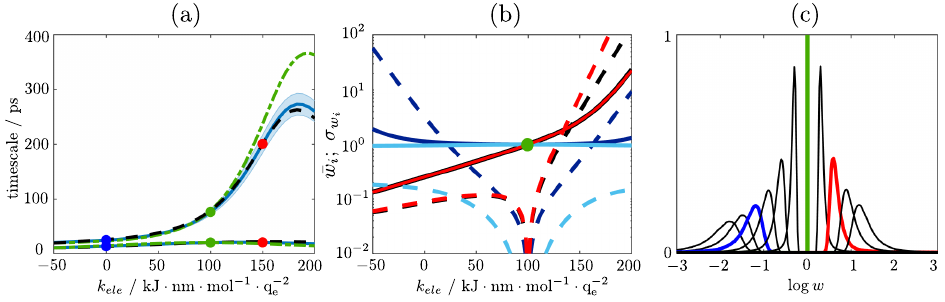}
\caption{
6-atoms molecule. 
(a) First two MSM (blue solid line) and SqRA (black and green dashed lines) implied timescales as functions of the electric constant;
(b) Average path-reweighting weights $\bar{w}_i$ (black solid line) and standard deviation $\sigma_{{w}_i}$ (black dashed line): $w_1$ (black), $w_2$ (dark blue), $w_3$ (light blue), $w$ (red);
(c) Distribution of the logarithm of the path-reweighting weights. 
The color denotes the electric constant: $k_{ele} =$ -50 (blue), 0 (green), 150 (red) $\mathrm{kJ \cdot nm \cdot mol^{-1} \cdot q_e^{-2}}$.
}
\label{fig:fig7}
\end{figure*}

\subparagraph{Implied timescales by Girsanov reweighting}
To study how the slowest implied timescales depend on $k_{ele}$, we constructed the MSM with a fixed lag time of 10 ps, for a continuous range of $k_{ele}$ values between -50 and $200\, \mathrm{kJ \cdot nm \cdot mol^{-1} \cdot q_e^{-2}}$.
The results are plotted in fig.~\ref{fig:fig7}-a (blue solid line).
If $k_{ele}< 50\, \mathrm{kJ \cdot nm \cdot mol^{-1} \cdot q_e^{-2}}$, then the two slowest implied timescales are almost constant ($\approx 15$ and 35 ps); if $k_{ele}> 50\, \mathrm{kJ \cdot nm \cdot mol^{-1} \cdot q_e^{-2}}$, then the two implied timescales diverge.
The electric constant $k_{ele} = 50\, \mathrm{kJ \cdot nm \cdot mol^{-1} \cdot q_e^{-2}}$ is indeed the threshold which determines the transition from a monostable to a bistable state system.
As the opening-closing of the molecule is a rare event, the first implied timescales quickly raises from $\approx 35 \, \mathrm{ps}$ to a maximum of $\approx 275 \, \mathrm{ps}$ at $k_{ele} = 180 \, \mathrm{kJ \cdot nm \cdot mol^{-1} \cdot q_e^{-2}}$.
Afterward, the first implied timescale falls again as the attractive force is so strong that the system is stable in the close conformation.
However, the path reweighting becomes unstable as suggested by the larger standard deviation.

To explain this observation, we analyzed the transition weights used to build the MSMs, which can be decoupled as
\begin{eqnarray}
w &=& M_{\tau,  r_{16}}(\omega) \cdot g(r_{16}) \cr
  &=& e^{-\beta U(r_{16})}\cdot e^{-I_S} \cdot e^{-\nicefrac{1}{2} I_R} \cr
  &=& w_1 \cdot w_2 \cdot w_3 \, ,
\label{eq:weight}
\end{eqnarray}
where $M_{\tau, r_{16}}(\omega)$ and $g(r_{16})$ are respectively the path-reweighting terms defined in eqs.~\ref{eq:girsanov}, \ref{eq:RadonNikodym01}, while $U(r_{16})$, $I_S$ and $I_R$ are defined respectively in eqs.~\ref{eq:Ur16}, \ref{eq:IS} and \ref{eq:IS}.
In fig.~\ref{fig:fig7}-b, we report the average weights $\bar{w}_1$, $\bar{w}_2$, $\bar{w}_3$ and $\bar{w}$ (black, dark blue, light blue and red solid line) and the standard deviations $\sigma_{w_1}$, $\sigma_{w_2}$, $\sigma_{w_3}$ and $\sigma_{w}$ (black, dark blue, light blue and red dashed line) as functions of $k_{ele}$, calculated over a single trajectory  with a lag time of 10 ps.
Note that $\bar{w}_1$ (black solid line) and $\bar{w}$ (red solid line) overlap.
The distributions of $\log w$ are reported in fig.~\ref{fig:fig7}-c, where each curve represents a distribution for a different $k_{ele}$ value.

We observe that if $k_{ele}^{\mathrm{target}}<100 \, \mathrm{kJ \cdot nm \cdot mol^{-1} \cdot q_e^{-2}}$, i.e. the value used to perform the direct simulation, the distribution of transition weights is bounded such that the standard deviation of the weights $\sigma_{w}$ is always smaller than the average weight $\bar{w}$.
In the neighborhood of $k_{ele}^{\mathrm{target}} = 100 \, \mathrm{kJ \cdot nm \cdot mol^{-1} \cdot q_e^{-2}}$,  $\bar{w} \approx 1$ and the standard deviation of the weights collapses to zero.
On the contrary, if $k_{ele}^{\mathrm{target}}> 100 \, \mathrm{kJ \cdot nm \cdot mol^{-1} \cdot q_e^{-2}}$,  $\bar{w} > 1$, and the distribution of the weights becomes quickly so broad that the standard deviation grows much faster than its average.
This results in transition probability matrices whose neighboring entries differ by several orders of magnitude, causing numerical instability when the eigenvectors and eigenvalues are computed.

The relationship between weights $w_i$ and $k_{ele}$ is due to the function $U(r) \sim \nicefrac{1}{r}$ in $w_1$, which dominates the terms $\nabla U(r) \sim - \nicefrac{1}{r^2}$ and $\left(\nabla U(r)\right)^2 \sim \nicefrac{1}{r^4}$ respectively in $w_2$ and $w_3$.
However, this is only true for small values of the lag time.
Indeed, the exponents in $w_2$ and $w_3$ are integrals over time, and for larger values of the lag time, they contribute more to the total weight $w$.
%

\subparagraph{Implied timescales by SqRA reweighting}
%
In contrast to the two-dimensional example, we applied the SqRA on a subspace of the state space, represented by the reaction coordinate $r_{16}$.
As we already discussed, this provides the correct left eigenvectors, but the eigenvalues are scaled up to the diffusion constant which appears in eq.~\ref{eq:flux02}.
Indeed, projecting the dynamics on reaction coordinates gives rise to a diffusion along the reaction coordinates which is unknown a priori.
Nonetheless, the ratio between SqRA eigenvalues is correct \cite{Lie2013}, then we determine the diffusion from the MSM implied timescales \cite{Donati2018b}.

As first attempt, we estimated the diffusion constant applying eq.~\ref{eq:Diffu} using the implied timescales obtained from the simulation run with $k_{ele} = 100\, \mathrm{kJ \cdot nm \cdot mol^{-1} \cdot q_e^{-2}}$,  obtaining the value
\begin{eqnarray}
\hat{D} = 0.032 \, \mathrm{nm^2 \cdot ps^{-1}} \, .
\end{eqnarray}
This provided the correct SqRA implied timescales (eq.~\ref{eq:tn}) in the range $k_{ele} \in [-50, 100] \,\mathrm{kJ \cdot nm \cdot mol^{-1} \cdot q_e^{-2}}$ as shown by the green dashed-dotted line in fig.~\ref{fig:fig7}-a.
Unfortunately, for $k_{ele} > 100 \,\mathrm{kJ \cdot nm \cdot mol^{-1} \cdot q_e^{-2}}$, the SqRA implied timescales are overestimated and the difference with respect to the MSM implied timescales exponentially grows: 
at $k_{ele} = 150 \,\mathrm{kJ \cdot nm \cdot mol^{-1} \cdot q_e^{-2}}$ the relative difference is 16\%;
at $k_{ele} = 180 \,\mathrm{kJ \cdot nm \cdot mol^{-1} \cdot q_e^{-2}}$ (at the maximum) the relative difference is 31\%.
The reason for this is that tuning the electric constant influences the diffusion along the reaction coordinate which is itself a function of $k_{ele}$.

\begin{figure*}[ht]
\includegraphics[]{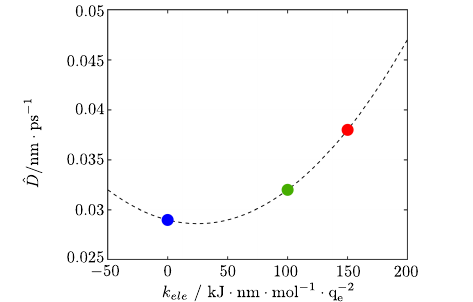}
\caption{
6-atoms molecule. Diffusion along the reaction coordinate estimated from the MSMs.
The color denotes the electric constant: $k_{ele} =$ -50 (blue), 0 (green), 150 (red) $\mathrm{kJ \cdot nm \cdot mol^{-1} \cdot q_e^{-2}}$.
}
\label{fig:fig8}
\end{figure*}

As second attempt, in order to improve our results we took advantage of the test simulations to estimate the diffusion constant also at $k_{ele} = 0$ and $150 \,\mathrm{kJ \cdot nm \cdot mol^{-1} \cdot q_e^{-2}}$.
As shown in fig.~\ref{fig:fig8} the diffusion constant slightly increases with the electric constant.
In order to get a continuous function of the diffusion constant, we connected the three diffusion values by a spline.
This was an arbitrary approximation, as we do not know how the diffusion depends on $k_{ele}$; on the other hand, we know that the implied timescales are only linearly sensitive to the diffusion, allowing a degree of discretion in the estimation of the diffusion.
This choice permitted to adjust the SqRA implied timescales which are in excellent agreement with those predicted by the Girsanov reweighting (fig.~\ref{fig:fig8}-a, black dashed line).
%

%
%
%
\section{Discussion and conclusion}
\label{sec:disc_sec}
We have presented the Girsanov reweighting method for MSMs \cite{Schuette2015b, Donati2017, Donati2018, Quer2018, Kieninger2020, Kieninger2021} and the SqRA method \cite{Lie2013, Donati2018b, Heida2018, Donati2021} to discretize the Fokker-Planck operator.

The first one is a dynamical reweighting method used to build the MSM of a target potential given a set of simulations performed at a reference potential.
Within the field of MD, the method is recommended to reweight simulations generated by enhanced sampling techniques, such as metadynamics \cite{Huber1994, Laio2002} and umbrella sampling \cite{Souaille2001}, to recover the dynamical information of the unbiased system from biased trajectories.
Here, after having reviewed the underlying theory, we proposed two different uses.
First, we employed Girsanov reweighting to study a convex combination of two potential energy functions, i.e. a transformation between two potentials through intermediates states tuned by an external parameter $\lambda$.
%
%
%
In the second example, we applied the method to investigate how the dynamics of a molecule depends on the Coulomb potential and the electric constant $k_{ele}$, showing that Girsanov reweighting is applicable in force field optimization.
For instance, one can combine path reweighting with the maximum caliber approach \cite{Bolhuis2021} to improve force field parameters such that the path probability density is minimally perturbed but the kinetics still match an external constraint \cite{Bolhuis2022}.

For both experiments, we applied the reweighting to trajectories generated at transitional potentials, i.e. defined by intermediate parameters $\lambda^{\mathrm{sim}}$ and $k_{ele}^{\mathrm{sim}}$, to predict the MSM eigenvectors, and the associated implied timescales, for any other parameter value in a reasonable range.
In principle, we could have generated trajectories from an \emph{extreme} potential, e.g. at $\lambda = 0$ or $\lambda = 1$ in the first example, and reweighted them to the opposite extreme potential, but this would have produced inaccurate results.
Indeed, not wrapping potentials do not satisfy the absolute continuity condition, the theoretical requisite required by the Girsanov theorem to guarantee the existence of the ratio between the associated probability densities (eq.~\ref{eq:absContPath}).

From a numerical standpoint, other precautions should be taken when dealing with the reweighting factor $M_{\tau,  x}(\omega)$ defined in eq.~\ref{eq:girsanov}, because it becomes quickly intractable if the paths $\omega$ are too long, or if the sum over the components of the gradient of the bias $\nabla U(x)$ is too large.
By using the Onsager-Machlup action to reweight time-correlation functions, Xing and Andricioaei \cite{Xing2006} have already observed this limitation; however, in MSMs, the lag-time can be chosen even several orders of magnitude smaller than the slowest process of the system, so that the Ito integral and Riemann integral in eq.~\ref{eq:girsanov} over short paths can be estimated.
Nonetheless, we remark that in MSMs a too short lag-time could cause the loss of Markovianity of the model, then the choice of the lag-time is not completely arbitrary \cite{Prinz2011}.
With regard the problem of the gradient, it is circumvented in our applications, including enhanced sampling MD simulations \cite{Donati2018b}, because the bias is formulated in terms of one or two collective variables and most elements of the gradient in eq.~\ref{eq:girsanov} are zero.
Hence, the sum over all the dimensions is actually estimated over a limited number of degrees of freedom.
In the second example, we also showed that the distribution of the weights affects the stability of the reweighing.
Long-tailed distributions imply reweighted transition probability matrices whose entries differ by several orders of magnitude.
This leads to inaccurate results and makes solving the Eigenvalue Problem difficult.
In most cases, such a situation occurs when the logarithm of the weights (eq.~\ref{eq:weight}) is positive, due to the exponential nature of the reweighting factors. 
Then, to optimize the reweighting, one should set the problem in order to have bounded exponents in the neighborhood of zero, in eqs.~\ref{eq:girsanov}, \ref{eq:RadonNikodym01}.

Another technical difficulty relates to when the reweighting factor should be calculated.
It is possible to save the random numbers used to solve eq.~\ref{eq:sde} and then calculate the gradient of the difference potential $U(x)$ during the post-analysis, but this approach would require considerable computational resources.
Instead, we recommend that the integrands of the reweighting factor are estimated on-the-fly, and the exponential function is calculated only during the construction of the MSM.
This can only be accomplished by modifying the MD program's source code. 
For example, OpenMM \cite{Eastman2013} provides a relatively straightforward way to implement this, and a number of applications were already realized in this way \cite{Donati2017, Donati2018, Kieninger2021}.
We finally remark that eq.~\ref{eq:girsanov} has been derived for overdamped Langevin dynamics applying the Euler-Maruyama scheme, but, it turned out to be an accurate approximation also for Langevin dynamics discretized with the leapfrog integrator implemented in OpenMM \cite{Izaguirre2010,Eastman2013}.
This was verified with numerical applications in refs.~ \cite{Donati2017, Donati2018}, and theoretically demonstrated by comparing eq.~\ref{eq:girsanov} with the exact reweighting formula for Langevin dynamics in ref.~\cite{Kieninger2021}.
However, integrators based on Strang splitting \cite{Kieninger2022} could require different path reweighting formulas.

The second method illustrated, the SqRA, can be used either as an alternative to MSMs or as a reweighting method, as shown respectively in our two examples.
The practical advantage is that it does not extract dynamical information from time-correlation functions of long trajectories, but from the stationary distribution of the system.
Since there are no integrals over time, SqRA results are numerically robust and their precision is solely determined by the granularity of the discretization \cite{Heida2018}.
Despite this, SqRA is limited to 9,10-dimensional systems \cite{Donati2018b}, while for higher-dimensional systems one needs to project the dynamics onto low-dimensional reaction coordinates.
This yields the correct eigenvectors in the low-dimensional reaction coordinate space, but at the expense of information about the eigenvalues, which can only be determined through MSMs methods.

In conclusion, Girsanov reweighting and SqRA are powerful tools designed to provide efficiently a Markovian representation of high-dimensional dynamical systems.
The former provides the correct eigenvectors and timescales, but it is susceptible to numerical problems; the latter is numerically robust, it provides the correct eigenvectors, but the timescales are not physically meaningful when working in a reduced space.
In the light of these considerations, we believe that these methods are a natural fit and that they can be combined into a unique reweighting scheme, matching the advantages of one with the limitations of the other.
%

\section*{Data availability statement}
The data that support the findings of this study are available from the corresponding author upon reasonable request.
\begin{acknowledgments}
This research has been funded by the Deutsche Forschungsgemeinschaft (DFG, German Research Foundation) through the grant CRC 1114 ``Scaling Cascades in Complex Systems'', project B05 ``Origin of the scaling cascades in protein dynamics'' and through grant SFB 1449 – 431232613, project C02; and the Cluster of Excellence MATH+, project AA1-15 ``Math-powered drug-design''.
\end{acknowledgments}
\appendix
\section{Derivation of eq.~\ref{eq:girsanov}}
\label{sec:appendix1}
Here, we recall the formal derivation of eq.~\ref{eq:girsanov}, i.e. the ratio between path probability densities as reported in \cite{Donati2017}.

Consider two independent one-dimensional stochastic differential equations 
\begin{eqnarray}
\mathrm{d}x_t &=& a(x_t) \, \mathrm{d}t + \sigma \mathrm{d}W_t \, ,\cr 
\mathrm{d}x_t &=& b(x_t) \, \mathrm{d}t + \sigma \mathrm{d}W_t \, ,
\label{eq:sde2} 
\end{eqnarray}
where $a(x)$ and $b(x)$ are the respective drift terms, $\sigma$ is a constant volatility, the same for both  equations, and $W_t$ is a Wiener process.
Applying the Euler-Maruyama scheme \cite{Leimkuhler2015} yields
\begin{eqnarray} 
x_{k+1} &=& x_{k} + a_k\Delta t + \zeta_k \sigma \sqrt{\Delta t} \, , \cr
x_{k+1} &=& x_{k} + b_k\Delta t + \eta_k \sigma \sqrt{\Delta t} \, .
\end{eqnarray}
where $\Delta t$ is the integration time step, $a_k = a(x_k)$ and $b_k = b(x_k)$, $\zeta_k$ and $\eta_k$ are two i.i.d random numbers drawn at timestep $k$ from a standard Gaussian distribution.
Given a particular path $\omega$ of time length $\tau = n \cdot \Delta t$ starting at $x_0=x \in \mathbb{R}$ generated by the first equation in eq.~\ref{eq:sde2}, the ratio between the path probability densities defined in eq.~\ref{eq:RadonNikodym} is written as 
\begin{eqnarray}
	M_{\tau, x}(\omega) &= &\frac{\mu_{P_b}(\omega)}{\mu_{P_a}(\omega)} 
=  \frac{\prod_{k=1}^n \exp \left(-\frac{(x_{k+1} - x_{k} - b_k\Delta t)^2}{2\Delta t \sigma^2} \right)}{\prod_{k=1}^n \exp \left(- \frac{(x_{k+1} - x_{k} - a_k\Delta t)^2}{2\Delta t \sigma^2} \right)} \, ,
\label{eq:appB3}
\end{eqnarray}
where $\mu_{P_a}$ and $\mu_{P_b}$ were defined in eq.~\ref{eq:path_prob_dens}.
Note that in eq.~(\ref{eq:appB3}), the normalization constants that appear in eq.~\ref{eq:ConditionalProbability} cancel.
Rearranging eq.~\ref{eq:appB3} yields
\begin{eqnarray}
	&&M_{\tau, x}(\omega) 
	= \cr
	&&\exp \left(\sum_{k=0}^n \frac{  (x_{k+1} - x_{k}) \left( b_k - a_k \right) }{\sigma^2} \right) \exp \left(-\sum_{k=0}^n \frac{ \left( b_k^2 - a_k^2 \right) \Delta t}{2 \sigma^2} \right) \, .
\label{eq:appB4}
\end{eqnarray}
Taking the limit $\Delta t \rightarrow 0$ the first term converges to the It\^{o} integral
\begin{equation}
\begin{aligned}
\lim_{\Delta t \rightarrow 0}  \sum_{k=0}^n (b_k-a_k)(x_{k+1} - x_{k}) = & \int_0^{\tau} (b(x_s)-a(x_s)) \mathrm{d}x_s \\ 
= & \int_0^{\tau} (b(x_s)-a(x_s))(a(x_s)\mathrm{d}s + \sigma \mathrm{d} W_s) \, ,
\label{eq:appB7}
\end{aligned}
\end{equation}
while the exponent in the second term of (\ref{eq:appB4}) converges to the Riemann integral
\begin{eqnarray}
	\lim_{\Delta t \rightarrow 0}\sum_{k=0}^n \frac{ \left( b_k^2 - a_k^2 \right) \Delta t}{2 \sigma^2}     
	&=& \frac{1}{2} \int_0^{\tau} \frac{b(x_s)^2 - a(x_s)^2}{\sigma^2} \mathrm{d} s \, .
\label{eq:appB5}
\end{eqnarray}
Inserting (\ref{eq:appB7})--(\ref{eq:appB5}) into equation (\ref{eq:appB4}) yields the Girsanov formula
\begin{eqnarray}
&&\lim_{\Delta t \rightarrow 0}  M_{\tau, x}(\omega) = \cr
  &&\exp \left(\int_0^{\tau} \frac{b(x_s) - a(x_s)} \sigma \mathrm{d} W_s \right) \exp \left(-\frac{1}{2} \int_0^{\tau} \frac{(b(x_s)-a(x_s))^2}{\sigma^2} \mathrm{d} s  \right) \, ,
\label{eq:appB8}
\end{eqnarray}
which expresses the ratio between path probability densities for time-continuous paths.
Applying the Euler-Maruyama scheme to the integrals, one obtains the discretized version reported in eq.~\ref{eq:girsanov}, which is used in practice applications.

\section{Force field parameters}
\label{sec:appendix2}
The parameters in tab.~\ref{tab:tab2} were used to simulate the 6-atoms molecule describe in the numerical experiments section.
%
%
\begin{table}[!ht]
\begin{tabular}{|cccc|}
\hline
\multicolumn{4}{|c|}{\textbf{Lennard-Jones potential}}                                                                                              \\ \hline
\multicolumn{1}{|c|}{Distance} & \multicolumn{1}{c|}{$\varepsilon_{ij}$ ($\mathrm{kJ \cdot mol^{-1} \cdot nm^{-2}}$)} & \multicolumn{1}{c|}{$r_{0,ij}$ (nm)}       &  \\ \cline{1-3}
\multicolumn{1}{|c|}{$r_{15}$}            & \multicolumn{1}{c|}{$0.5$}
& \multicolumn{1}{c|}{0.5}              &                   \\ \cline{1-3}
\multicolumn{1}{|c|}{$r_{16}$}            & \multicolumn{1}{c|}{0.5}              & \multicolumn{1}{c|}{1.0}                 &                   \\ \cline{1-3}
\multicolumn{1}{|c|}{$r_{26}$}            & \multicolumn{1}{c|}{$0.5$}             & \multicolumn{1}{c|}{1.5}                 &                   \\ \hline\hline
\multicolumn{4}{|c|}{\textbf{Bond distance potential}}                                                                                              \\ \hline
\multicolumn{1}{|c|}{Bond distance}       & \multicolumn{1}{c|}{$k_{ij}$ ($\mathrm{kJ \cdot mol^{-1} \cdot nm^{-2}}$)}      & \multicolumn{1}{c|}{$r_{0,ij}$ (nm)}       & \\ \cline{1-3}
\multicolumn{1}{|c|}{$r_{12}$}            & \multicolumn{1}{c|}{$50$}              & \multicolumn{1}{c|}{1.69}                 &                   \\ \cline{1-3}
\multicolumn{1}{|c|}{$r_{23}$}            & \multicolumn{1}{c|}{50}              & \multicolumn{1}{c|}{1.50}                 &                   \\ \cline{1-3}
\multicolumn{1}{|c|}{$r_{34}$}            & \multicolumn{1}{c|}{50}              & \multicolumn{1}{c|}{1.79}                 &                   \\ \cline{1-3}
\multicolumn{1}{|c|}{$r_{45}$}            & \multicolumn{1}{c|}{50}              & \multicolumn{1}{c|}{1.56}                 &                   \\ \cline{1-3}
\multicolumn{1}{|c|}{$r_{56}$}            & \multicolumn{1}{c|}{$50$}              & \multicolumn{1}{c|}{1.54}                 &                   \\ \hline\hline
\multicolumn{4}{|c|}{\textbf{Angle potential}}                                                                                                      \\ \hline
\multicolumn{1}{|c|}{Angle}               & \multicolumn{1}{c|}{$k_{ijk}$ ($\mathrm{kJ \cdot mol^{-1} \cdot rad^{-2}}$)}     & \multicolumn{1}{c|}{$\theta_{0,ijk}$ (rad)} &  \\ \cline{1-3}
\multicolumn{1}{|c|}{$\theta_{123}$}      & \multicolumn{1}{c|}{$50$}              & \multicolumn{1}{c|}{$\frac{2}{3}\pi$}                 &                   \\ \cline{1-3}
\multicolumn{1}{|c|}{$\theta_{234}$}      & \multicolumn{1}{c|}{50}              & \multicolumn{1}{c|}{$\frac{2}{3}\pi$}                 &                   \\ \cline{1-3}
\multicolumn{1}{|c|}{$\theta_{345}$}      & \multicolumn{1}{c|}{50}              & \multicolumn{1}{c|}{$\frac{2}{3}\pi$}                 &                   \\ \cline{1-3}
\multicolumn{1}{|c|}{$\theta_{456}$}      & \multicolumn{1}{c|}{$50$}              & \multicolumn{1}{c|}{$\frac{2}{3}\pi$}                 &                   \\ \hline\hline
\multicolumn{4}{|c|}{\textbf{Torsion angle potential}}                                                                                                    \\ \hline
\multicolumn{1}{|c|}{Torsion angle}       & \multicolumn{1}{c|}{$k_{ijkl}$ ($\mathrm{kJ \cdot mol^{-1} \cdot rad^{-2}}$)}    & \multicolumn{1}{c|}{$m_{ijkl}$}       & $\Psi_{0,ijkl} (\mathrm{rad})$   \\ \hline
\multicolumn{1}{|c|}{$\Psi_{1234}$}       & \multicolumn{1}{c|}{$20$}              & \multicolumn{1}{c|}{1}                 &        $\pi$           \\ \hline
\multicolumn{1}{|c|}{$\Psi_{2345}$}       & \multicolumn{1}{c|}{20}              & \multicolumn{1}{c|}{1}                 &         $\frac{1}{6}\pi$       \\ \hline
\multicolumn{1}{|c|}{$\Psi_{3456}$}       & \multicolumn{1}{c|}{$20$}              & \multicolumn{1}{c|}{1}                 &         $\pi$          \\ \hline\hline
\end{tabular}
\caption{Force field parameters for the 6-atom molecule.}
\label{tab:tab2}
\end{table}
%


\bibliographystyle{unsrt}
\bibliography{references.bib}

\end{document}